\documentclass[11pt,fleqn]{book}

\usepackage{setspace}
\usepackage{simplethesis}
\usepackage{amsmath,amssymb,amsthm}
\usepackage{proof}
\usepackage[all]{xy}
\usepackage[pdftitle={The Suspension Calculus and its Relationship to
  Other Explicit Treatments of Substitution in Lambda Calculi},
  pdfauthor={Andrew Gacek},
  colorlinks=true,
  linkcolor=blue,
  citecolor=blue,
  urlcolor=blue,
  plainpages=false]{hyperref}

\long\def\ignore#1{}

\newcommand{\ra}{\rightarrow}

\newcommand{\app}{{\ }}

\newcommand{\lambdadb}{\lambda \,}
\newcommand{\lambdadbann}[1]{\lambda_{#1} \,}

\newcommand{\onebeta}{\rhd_{\!\beta}}
\newcommand{\mbeta}{\rhd_{\!\beta}^*}

\newcommand{\one}[1]{\rhd_{\!#1}}
\newcommand{\oneread}{{\rhd\!_{r}}}
\newcommand{\onemerge}{{\rhd\!_{m}}}
\newcommand{\onebetas}{{\rhd\!_{\beta_s}}}
\newcommand{\onereadmerge}{{\rhd\!_{rm}}}
\newcommand{\onereadbetas}{{\rhd\!_{r\beta_s}}}
\newcommand{\oneall}{\rhd\!_{rm \beta_s}}

\newcommand{\many}[1]{\rhd_{\!#1}^*}

\newcommand{\merge}{{\rhd\!_{m}^*}}
\newcommand{\readmerge}{{\rhd\!_{rm}^*}}
\newcommand{\readbetas}{{\rhd\!_{r\beta_s}^* }}
\newcommand{\betas}{{\rhd\!_{\beta_s}^*}}
\newcommand{\all}{\rhd\!_{rm \beta_s}^*}

\newcommand{\posread}{{\rhd\!_{r}^+}}

\newcommand{\onereadmergep}{{\rhd\!_{rm'}}}

\newcommand{\rnf}[1]{{\vert #1 \vert}}

\newcommand{\lenv}{{\lbrack\!\lbrack}}
\newcommand{\renv}{{\rbrack\!\rbrack}}
\newcommand{\env}[1]{{\lenv #1 \renv}}

\newcommand{\lmenv}{{\lbrace\!\!\lbrace}}
\newcommand{\rmenv}{{\rbrace\!\!\rbrace}}
\newcommand{\menv}[1]{{\lmenv #1 \rmenv}}

\newcommand{\lmenvt}{{\langle\!\langle}}
\newcommand{\rmenvt}{{\rangle\!\rangle}}
\newcommand{\menvt}[1]{{\lmenvt #1 \rmenvt}}

\newcommand{\monussign}{{\stackrel{.}{\;\overline{\,\,\,}\;}}}
\newcommand{\monus}[2]{{#1 \monussign #2}}

\newcommand{\ie}{{\em i.e.}}
\newcommand{\eg}{{\em e.g.}}

\newcommand{\yields}{\vdash_\Sigma}

\newcommand{\ess}{{\cal E}}

\newcommand{\Thechapter}{\thechapter}

\newcommand{\lift}{\mathop{\Uparrow}}

\newcommand{\sig}[1]{\mathop{\sigma^{#1}}}
\newcommand{\ph}[2]{\mathop{\varphi_{#1}^{#2}}}

\newtheorem{defn}{Definition}[section]
\newtheorem{theorem}{Theorem}[section]
\newtheorem{lemma}{Lemma}[section]

\begin{document}

\Author{Andrew Jude Gacek}
\Title{The Suspension Calculus and its Relationship to Other Explicit
  Treatments of Substitution in Lambda Calculi}
\Month{December}
\Year{2006}
\Adviser{Gopalan Nadathur}
\Degree{MASTER OF SCIENCE}
\degreelevel{masters}

\prelimpages

\titlepage

\copyrightpage

\setcounter{page}{1}

\acknowledgments{Many people have supported me during this thesis and I owe them all a
debt of gratitude. In particular, I would like to first thank my wife
Ann for keeping my spirits up and my nose to the grindstone.  I would
like to thank Gopalan Nadathur, my advisor, for constantly sharing his
knowledge and needed criticism. In addition, I would like to thank
Dale Miller, Wayne Richter, and Ravi Janardan for serving on my
committee and allowing me to present my ideas to fresh ears and open
minds. Finally I would like to thank my good friend Jared Davis for
explanations, discussions, and advice.

This thesis is based on work supported by the National Science
Foundation under Grant 0429572 and also by Boston Scientific.  Any
opinions, findings, and conclusions or recommendations expressed in
this material are those of the author and do not necessarily reflect
the views of the National Science Foundation or Boston Scientific.

}

\abstract{The intrinsic treatment of binding in the lambda calculus makes it an
ideal data structure for representing syntactic objects with binding
such as formulas, proofs, types, and programs. Supporting such a data
structure in an implementation is made difficult by the complexity of
the substitution operation relative to lambda terms. To remedy this,
researchers have suggested representing the meta level substitution
operation explicitly in a refined treatment of the lambda calculus.
The benefit of an explicit representation is that it allows for a
fine-grained control over the substitution process, leading also to
the ability to intermingle substitution with other operations on
lambda terms. This insight has lead to the development of various
explicit substitution calculi and to their exploitation in new
algorithms for operations such as higher-order unification.
Considerable care is needed, however, in designing explicit
substitution calculi since within them the usually implicit operations
related to substitution can interact in unexpected ways with notions
of reduction from standard treatments of the lambda calculus.

This thesis describes a particular realization of explicit
substitutions known as the suspension calculus and shows that it has
many properties that are useful in a computational setting. One
significant property is the ability to combine substitutions. An
earlier version of the suspension calculus has such an ability, but
the complexity of the machinery realizing it in a complete form has
deterred its direct use in implementations. To overcome this drawback
a derived version of the calculus had been developed and used in
practice.  Unfortunately, the derived calculus sacrifices generality
and loses a property that is important for new approaches to
unification.  This thesis redresses this situation by presenting a
modified form of the substitution combination mechanism that retains
the generality and the computational properties of the original
calculus while being simple enough to use directly in implementations.
These modifications also rationalize the structure of the calculus,
making it possible to easily superimpose additional logical structure
over it. We illustrate this capability by showing how typing in the
lambda calculus can be treated in the resulting framework and by
presenting a natural translation into the $\lambda\sigma$-calculus,
another well-known treatment of explicit substitutions.

Another contribution of this thesis is a survey of the realm of
explicit substitution calculi. In particular, we describe the
computational properties that are desired in this setting and then
characterize various calculi based on how well they capture these. We
utilize the simplified suspension calculus in this process.  In
particular, we describe translations between the other popular calculi
and the suspension calculus towards understanding and contrasting
their relative capabilities. Finally, we discuss an elusive property
of explicit substitution calculi known as preservation of strong
normalization and discuss why there is hope that the suspension
calculus possesses this property.

}

\setcounter{tocdepth}{2}
\tableofcontents

\textpages

\onehalfspacing

\chapter{Introduction}

Binding and scoping of variable names is a fundamental concept in
mathematics and computer science. Consider the mathematical
statement that the natural numbers have no largest element,
\[ \forall x\in \mathbb{N}.\ \exists y\in \mathbb{N}.\ y > x \]
In this statement the occurrences of $x$ and $y$ on the right are bound
by the quantifiers on the left, and the scoping of these variables is
important. Picking $y$ as the value for $x$ and substituting
blindly would yield,
\[ \exists y\in \mathbb{N}.\ y > y \] But this statement is no longer
true because the substitution was not done correctly. The problem is
that $y$ is not yet in scope when the quantifier for $x$ appears, thus
any substitution for $x$ cannot contain the variable $y$. This
situation is replayed in the following computer science context:
\begin{verbatim}
     int x = 3;
     int y = x + 2;
\end{verbatim}
Here the variables {\tt x} and {\tt y} are bound by declaration of
{\tt int x} and {\tt int y}, and here again the scoping is important:
we can use {\tt x} in defining {\tt y} but not vice-versa. The point
of these examples is that if we are going to represent and manipulate
objects from mathematics and computer science, then we should use a
language which has an understanding of binding and scoping.

\section{The Lambda Calculus and Its Treatment of Binding}

The lambda calculus is a notation for functions which correctly
captures notions of binding and scoping \cite{church-40-types}. For
example, the function which maps $x$ to $x + x$ is encoded as
$(\lambda x. x + x)$, and applying this function to the argument 3 is
encoded as $((\lambda x . x + x)\app 3)$. We expect that this
expression is equal to $3 + 3$, and indeed the lambda calculus
formalizes this notion of equality so that the equation $((\lambda x .
x + x)\app 3) = 3 + 3$ holds. Binding in the lambda calculus is done
such that variable occurrences are bound by the closest enclosing
binder which matches the variable name. For instance, in $(\lambda x .
\lambda x .  x + x)$ the two occurrences of $x$ on the right are bound
by second lambda.  Substitution in the lambda calculus is capture
avoiding in the sense that a free variable, one which is not bound by
a lambda, cannot become bound by the process of substitution. Thus it
is {\it not} true that $((\lambda x . \lambda y .  x)\app y) =
(\lambda y . y)$.  Instead, the bound variable $y$ is first renamed
before the free variable $y$ is substituted which yields the true
equality $((\lambda x. \lambda y.  x)\app y) = (\lambda
z. y)$. Finally note that we could have picked a name other than $z$
here so long as it did not capture the free variable $y$, because
renaming of bound variables is an intrinsic property of equality in
the lambda calculus.

Given this notion of equality, we can think of assigning
directionality to it so that $((\lambda x . x + x)\app 3) \ra 3 + 3$.
This directionality gives rise to a notion of computation through
function evaluation, which is common to most programming languages.
Based on this idea, Landin argues that we can understand any
particular programming language by translating it into the lambda
calculus \cite{landin-66-next700}. Such a translation actually turns
out to be a useful means for providing denotational semantics for
programming languages based on the model theory for the lambda
calculus developed by Plotkin, Scott, and Strachey
\cite{stoy-81-denotational}. Another benefit of thinking of the lambda
calculus as a common substrate for programming languages is that the
essence of issues such as typing and evaluation strategies can be
studied without the distraction of auxiliary features and nuances of
any particular programming language.

A more recent development related to the lambda calculus is its use in
representing syntactic objects whose structure incorporates some form
of binding. The motivation behind this is that common notions of
binding in these settings can be handled using the lambda calculus as
a data structure. For example, the content of quantifiers such as
``for all'' and ``there exists'' in mathematical logic can be
separated into two parts:  one part which is the binding of a variable
and another part which is 
predicative in nature. The predicative part can be represented by an
appropriate constant such as $forall$ or $exists$ and the binding part
can be represented by a lambda abstraction. Using these ideas, the
formula $\forall x\exists y(y > x)$ considered earlier in this
chapter can be represented by the lambda calculus expression
\[ forall\ (\lambda x.\ exists\ (\lambda y.\ gt\ y\ x)) \] where $gt$
is a constant representing the ``greater than'' relation. Using this
representation, logical operations related to binding can be performed
using the lambda calculus. For example, if we want to substitute $y$
for $x$ in the equation
\[ \forall x\in \mathbb{N}.\ \exists y\in \mathbb{N}.\ y > x \]
We do this by applying the argument of our $forall$ constant to the
variable $y$ and then performing a reduction in the lambda calculus
\[ (\lambda x.\ exists\ (\lambda y.\ gt\ y\ x))\app y \ra exists\
(\lambda z.\ gt\ z\ y) \]
This result corresponds to the logical formula
\[ \exists z\in \mathbb{N}.\ z > y \]
Similarity we may wish to identify our original statement with the
statement
\[ \forall a\in \mathbb{N}.\ \exists b\in \mathbb{N}.\ b > a \]
and this is provided for free using lambda calculus.

Similar notions of binding and substitution occur when representing
many different kinds of objects such as formulas, proofs, types, and
programs. The idea is that we can use a single meta language, the
lambda calculus, to talk about all of these different objects. Then
the correctness concerns of binding and substitution only need to be
handled once, in this meta language, and then the benefit is shared in
all other contexts.

\section{The Explicit Treatment of Substitution}

At the heart of the lambda calculus is a notion of substitution which
respects the binding structure of functions in the language. The
traditional presentation of the lambda calculus takes this
substitution as a meta operation, which is impractical from an
implementation perspective. Because of this, various methods have been
developed for dealing with substitution in actual implementations. In
the computational setting, what is often done is that an environment
is kept which contains substitutions for variables.  This approach has
been successful in practice, but it restricts the possible evaluation
strategies, since it expects that every variable encountered has a
substitution available in the environment. For example, this
expectation will not hold if we do evaluations underneath lambda
abstractions since the abstracted variables have no substitutions.

In the representational setting, substitution is also a problem. Here
we often perform unification modulo the rules of the lambda calculus,
and the handling of substitution has a significant impact on the
efficiency of this operation. Consider, for example, the unification
$(\lambda x.  c\app t_1)\app t_2 \stackrel{?}{=} d\app t_3$ for
distinct constants $c$ and $d$ and some terms $t_1$, $t_2$, and $t_3$.
A naive approach requires substituting the term $t_2$ for $x$
throughout the the term $t_1$, which can be arbitrarily large.
Instead, a more sophisticated technique is to treat substitution
explicitly and include it as part of the language. Then we can reduce
this unification problem to $c\app t_1 \langle t_2/x \rangle
\stackrel{?}{=} d\app t_3$ where $\langle t_2/x \rangle$ encodes a
substitution that is part of the language and not a meta operation. At
this point we can answer ``no'' since $c$ and $d$ are distinct
constants, and thus we avoid traversing the term $t_1$.

Explicit treatment of substitution introduces its own difficulties.
One particular difficulty stems from our interest in identifying
expressions that differ only in the names of bound variables. To
accommodate this, we use a nameless notation for the lambda calculus
invented by de Bruijn \cite{debruijn-72-nameless}. In this notation,
we replace each variable occurrence with an integer which counts the
number of lambda abstractions between the occurrence and its binder.
For example, the lambda term $(\lambda x. \lambda y. x\app x\app y)$
is encoded as $(\lambdadb \lambdadb 2\app 2\app 1)$. This provides a
unique encoding of terms differing only in bound variable renaming,
but now substitution must take into account renumberings when
substituting underneath a lambda abstraction. For example,
$(\lambdadb\lambdadb ((\lambdadb \lambdadb 2)\app 2)) =
(\lambdadb\lambdadb\lambdadb 3)$.  Thus any system for explicit
substitutions must keep track of the renumbering to be done.

The explicit treatment of substitutions also gives rise to new
benefits in the lambda calculus. One is that we may think of merging
substitutions to decrease the amount of work done in traversing a
term. For example, reducing $((\lambda x. \lambda y. t_1)\app
t_2\app t_3)$ traditionally requires two traversals over the term
$t_1$, but if the substitutions generated by reducing the above term
are explicit then we can think of merging them into a single
substitution before their effects are propagated onto $t_1$. Another
benefit is that existing procedures such as unification can be
improved upon by mixing their operations with the operations of
explicit substitutions \cite{dowek-95-higherorder}.

These motivations for the explicit treatment of substitutions have
been recognized by researchers, resulting in a wide variety of
explicit substitution calculi \cite{field-90-laziness,
  nadathur-90-representation, abadi-91-explicit,
  kamareddine-95-calculus, benaissa-96-upsilon,
  kamareddine-97-extending, david-01-calculus}. This thesis
focuses on the suspension calculus \cite{nadathur-90-representation}.

\section{Contributions of the Thesis}

In this thesis we use the suspension calculus as a viewpoint into the
realm of explicit substitution calculi. The particular contributions
of this thesis are

\bigskip

\noindent {\bf A simplification and rationalization of the suspension
  calculus}\\
The suspension calculus includes not only an explicit representation
of substitutions, but also a mechanism for combining such
substitutions. This combining is realized through a seemingly complex
set of rules, and this apparent complexity has lead to the development
of derived calculi \cite{nadathur-99-finegrained}. These derived
calculi work well for reduction, but they lack an essential property
and therefore they cannot be used to perform unification in a setting
with a special interpretation of instantiable variables. This thesis
offers another possibility by simplifying these combination rules so
that the full calculus is practical to use in implementations, while
retaining all of the essential properties of the original suspension
calculus. These changes have the added benefit of rationalizing the
calculus so that a typed version of the calculus is possible, and
translations to other calculi are now feasible.

\bigskip

\noindent {\bf A comparison of explicit substitution calculi}\\
A wide variety of explicit substitution calculi have been proposed,
but no systematic attempt has been made to compare these calculi and
analyze their essential features. In this thesis we outline desirable
properties for explicit substitution calculi and use these to organize
a survey of the more popular calculi. We also explore the relationship
between these calculi and the suspension calculus by defining explicit
translations for expressions. These translations serve as a means of
understanding the notation and rules of each calculus using the
framework of the suspension calculus.

\section{Outline of the Thesis}

The rest of this thesis is organized as follows. In the next chapter
we review the lambda calculus and introduce the terminology associated
with it that we will use in later chapters. We then describe the
suspension calculus in \autoref{ch:susp} and prove its key properties.
In \autoref{ch:compare} we survey other explicit substitution calculi
and compare them with the suspension calculus. Then in
\autoref{ch:psn} we discuss the property of preservation of strong
normalization, which is an important issue for explicit substitution
calculi. Finally, in \autoref{ch:conc} we review the contributions of
this thesis and discuss possible avenues for future research.


\chapter{The Lambda Calculus}

The lambda calculus is a language of functions---a simple and concise
syntax for describing a powerful and expressive language. As discussed
in the introduction, this calculus has two related uses at a practical
level: (1) to represent syntactic objects that have a functional
structure in a way that captures their functionality and (2) by
interpreting representations of functions essentially as rules for
calculations, it also supports the ability to encode computations in a
fundamental way. In short, we call these two uses ``representation''
and ``computation,'' respectively.  The goal of this chapter is to
define the calculus and present notation and properties that support
these two different uses.

In the first section, we define the lambda calculus and various
notions of equality on lambda terms. Following that, we look at the de
Bruijn notation for the lambda calculus which precisely captures the
most fundamental of these equality notions. In the third section, we
look at properties of the lambda calculus which are important to its
consistent and meaningful use in the roles of representation and
computation. Finally, we look at adding instantiable variables to
the calculus to support higher-order unification in the
representational setting. The results presented in this chapter are
well established in the literature and proven, for instance, in
\cite{barendregt-81-lambda}.

\section{The Syntax and Meaning of Lambda Terms}

Is $x + x$ a function? The answer depends on the context. In an
equation such as $x + x = 4$, we think of $x + x$ as a value, not a
function. On the other hand, in the statement, ``$x+x$ is primitive
recursive,'' we are thinking of the function which maps $x$ to $x +
x$. Church resolved this ambiguity by introducing an explicit notation
for functions, the lambda calculus \cite{church-40-types}. In the
lambda calculus we denote a function mapping $x$ to $x + x$ by
$(\lambda x . x + x)$. The lambda in this expression creates an
abstraction over $x$ so that $x$ is a bound variable within the
subexpression $x + x$. Juxtaposition denotes application, and so
$((\lambda x . x + x)\app 5)$ is the application of our function to
the argument 5. Later in this section we will introduce notions of
equality that capture our intuition about function equality and
evaluation. Surprisingly, abstraction and application along with these
notions of equality can model all computable functions.

With the syntax above, we can now use functions in a first-class way.
That is, we can use functions not only to manipulate input and produce
output, but also as the input and output of other functions. Consider
the function $(\lambda x . (\lambda y . x + y))$. This is a function
which when applied to some input, produces another function as output.
This allows us to encode functions of more than one argument by
nesting functions of exactly one argument. As another example,
consider $(\lambda f . (\lambda x . f\app x))$. This function takes a
function $f$ and an argument $x$ and returns the application of $f$ to
$x$. In this instance, we are thinking of a function as input to
another function. In this way, functions in the lambda calculus are
first-class and higher-order---an expressive notion captured by simple
syntax.

\subsection{Syntax}

Formally, the terms in the lambda calculus are defined by the
following.
\begin{defn}[Lambda terms]
\label{def:lambda-terms} Terms in the lambda calculus are defined
by
\[
t ::= c \ |\ x\ |\ (t\app t)\ |\ (\lambda x. t)
\]
where $c$ ranges over some enumerable set of constants and $x$ over
some enumerable set of variables.
\end{defn}

We call $(t\app t)$ an application and $(\lambda x . t)$ an
abstraction.  To differentiate between variables and constants, we
denote constants by letters like $a,b,c$ or by appropriate symbols,
and we denote variables by letters like $x, y, z$. To reduce the
number of parenthesis we need to write, we follow the convention that
application is left associative and the scope of lambda extends as far
to the right as possible. For example, $(\lambda x .  ((x\app y)\app
z))$ is written as $(\lambda x . x\app y\app z)$. We will sometimes
include optional parenthesis when they aid readability.

The syntax in \autoref{def:lambda-terms} is a solid starting point for
the lambda calculus, but there are many notions we need to build on
top of this. First, the definition of syntax yields an obvious
definition of subterm which we will assume.  Second, the syntax of the
lambda calculus contains a notion of binding which we must define
explicitly. This is crucial because the binding structure of a term is
its most essential part. All variable occurrences within a term are
either free or bound. The basic rule of binding is that a variable
occurrence of $x$ is free if and only if it does not occur within the
scope of a $\lambda x$, and all free occurrences of the variable $x$
in the term $t$ are bound by the lambda in $\lambda x . t$.

When picking names for bound variables, we must be careful that they
do not conflict with the free variables in a term. Thus the following
definition will be very useful.
\begin{defn}[Free variables]
\label{def:free-variables} The set of free variables of a lambda
term $t$, denoted $fv(t)$ is defined recursively as follows,
\begin{align*}
fv(x) &= \{x\} \\
fv(\lambda x . t) &= fv(t) \backslash \{x\} \\
fv(t_1\ t_2) &= fv(t_1) \cup fv(t_2)
\end{align*}
If $fv(t) = \emptyset$ then $t$ is called a closed term.
\end{defn}

Now that we have a clear notion of the binding structure of terms, we
can relate terms based on this structure. In the next section, we look
at notions of equality based on the binding structure of terms.

\subsection{Equality and Equivalence}

The introduction of the lambda calculus as a language of functions
gives us some preconceived notions of how lambda terms are
related. One is that we expect the specific variable names used for
bound variables to be irrelevant. Concretely, we consider the terms
$(\lambda x .  x + x)$ and $(\lambda y . y + y)$ equal, because they
represent the same function.

We also expect a notion of equality under function evaluation. So the
terms $((\lambda x . x + x)\ 5)$ and $5 + 5$ should be equal in this
sense. Informally, we want to say that the term $(\lambda x. t_1)\
t_2$
is equal to the term $t_1[t_2/x]$ where $[t_2/x]$ is an operator
which replaces all free occurrences of $x$ in $t_1$ with $t_2$. This
substitution operator must respect the binding structure of both $t_1$
and $t_2$. Specifically, we respect the structure of $t_1$ by not
substituting $t_2$ in for any bound occurrences $x$, and we respect the
structure of $t_2$ by not allowing any free variables in $t_2$ to
become bound in $t_1$. These restrictions are what gives rise to the
various branching conditions in the following definition.

\begin{defn}[Substitution]
\label{def:substitution} The substitution operation $[s/x]$ which
replaces the variable $x$ with the term $s$ is defined recursively
as
\begin{align*}
c[s/x] &= c \\
y[s/x] &= \begin{cases} s & \text{if $x = y$} \\
y & \text{otherwise}
\end{cases} \\
(t_1\ t_2)[s/x] &= (t_1[s/x])\ (t_2[s/x]) \\
(\lambda y . t) [s/x] &= \begin{cases}
\lambda y . t & \text{if $x = y$}\\
\lambda y' . (t [y'/y] [s/x]) &
\text{if $y \in fv(s)$, where $y' \notin fv(y)$ and $y' \neq x$} \\
\lambda y . (t[s/x]) & \text{otherwise}
\end{cases}
\end{align*}
\end{defn}

The middle case for $(\lambda y . t) [s/x]$ takes advantage of our
notion of equality for terms that differ only in the names of bound
variables. In this case, we do not want a free occurrence of $y$ in $s$
to be captured by the binding lambda, so we rename the bound variable
before performing the substitution.

With a definition of substitution in place, we can go back and
formally define what we mean by terms that differ only in the names
of bound variables.

\begin{defn}[$\alpha$-equivalence]
\label{def:alpha-equivalence} A term $s$ results from a term $r$ by
$\alpha$-conversion if $s$ can be obtained from $r$ by replacing some
subterm of the form $\lambda x . t$ by one of the form $\lambda y .
(t[y/x])$ where $y$ is a variable that is not free in $t$.  Two terms
$s$ and $r$ are said to be $\alpha$-equivalent, written as $s =_\alpha
r$, if one can be obtained from the other by a (possibly empty)
sequence of $\alpha$-conversions.
\end{defn}

We can also formalize a notion of equivalence under evaluation that we
suggested before.

\begin{defn}[$\beta$-equivalence]
\label{defn:beta-equivalence}
A term $s$ results from a term $r$ by a $\beta$-contraction, denoted
$r \onebeta s$, if $s$ can be obtained by replacing a subterm of $r$
of the form $(\lambda x .  t_1)\ t_2$, referred to as a $\beta$-redex,
by $t_1[t_2/x]$. We say also that $s$ results from $r$ by a
$\beta$-reduction, denoted $r \mbeta s$, if it can be obtained from
$r$ by a (possibly empty) sequence of $\alpha$-conversions and
$\beta$-contractions. The term $r$ is said to result from $s$ by a
$\beta$-expansion if $s$ results from $r$ by a $\beta$-contraction.
Finally, $r$ and $s$ are said to be $\beta$-equivalent of one results
from the other by a sequence of $\alpha$-conversions,
$\beta$-contractions, and $\beta$-expansions, and we denote this by $r
=_\beta s$.
\end{defn}

Note that $\beta$-contraction invokes substitution which may cause the
renaming of some bound variables. For this reason we include
$\alpha$-conversion in our definition of $\beta$-reduction and
$\beta$-equivalence.

A final notion of equality that we might expect from functions is that
of extensional equivalence. That is, given $r\ x =_\beta s\ x$ we
might expect a notion of equality that says $r$ and $s$ are
equal. First note that $\beta$-equivalence is not powerful enough for
this. For instance, the terms $(\lambda x . f\app x)$ and $f$ are
related in this way, but they are not $\beta$-equivalent. It turns out
that we can get this extensionality property through the following
equivalence notion.

\begin{defn}[$\eta$-equivalence]
\label{def:eta-conversion}
A term $s$ results from a term $r$ by a $\eta$-contraction if $s$ can
be obtained by replacing a subterm of $r$ of the form $(\lambda x .
f\app x)$, referred to as a $\eta$-redex, by $f$, where $x$ is not
free in $f$. The term $r$ is said to result from $s$ by a
$\eta$-expansion if $s$ results from $r$ by a $\eta$-contraction.
Finally, $r$ and $s$ are said to be $\eta$-equivalent of one results
from the other by a sequence of $\alpha$-conversions,
$\beta$-conversions, $\eta$-contractions, and $\eta$-expansions, and
we denote this by $r =_\eta s$.
\end{defn}

Throughout this section we have referred to our equality notions as
equivalence relations, and the following theorem justifies this.

\begin{theorem}
$=_\alpha$, $=_\beta$, and $=_\eta$ are equivalence relations.
\end{theorem}

\section{De Bruijn Notation}

Substitution in the lambda calculus is complicated by the possibility
of variable names conflicting. De Bruijn notation is a nameless
notation for the lambda calculus which abstracts away many of these
issues \cite{debruijn-72-nameless}. The purpose of variable names in
the lambda calculus is to associate variable occurrences with their
binder. The de Bruijn notation makes this association by counting the
number of lambdas that occur between a variable occurrence and its
binder in the abstract syntax tree. In this way, names are removed
both from variable occurrences and from binders. For example, the term
$(\lambda x .  (\lambda y . y)\app (\lambda z . x))$ is encoded as
$(\lambdadb (\lambdadb \#1)\app (\lambdadb \#2))$.  This term has the
same content as the original, but we have abstracted away the specific
variable names.

\subsection{Terms in the De Bruijn Notation}

Formally, the terms in the lambda calculus are defined by the
following.

\begin{defn}[De Bruijn terms]
\label{def:debruijn-terms} Terms in the de Bruijn notation are
defined by
\[
t ::= c\ |\ \#i\ |\ (t\app t)\ |\ (\lambdadb t)
\]
where $c$ ranges over an enumerable set of constants and $i$,
called an index or variable reference, ranges over the
natural numbers.
\end{defn}

As with the definition of lambda terms, we call $(t\app t)$ an
application and $(\lambdadb t)$ an abstraction. We also drop
parenthesis by assuming application is left associative and the scope
of a lambda extends as far right as possible. In addition, we assume
the obvious definition of subterm.

One issue which we have not yet clarified is how to deal with free
variables. Because free variables have no binders associated with
them, it is not obvious how to assign them an index. To handle this we
will assume a fixed listing of the free variables of a term, which we
think of as a list of top level binders for the free variables. Thus
the term $(\lambda x . y\app x)$ where $y$ is free will be encoded as
$(\lambdadb \#2\app \#1)$. In this term, the $\#2$ refers to the first
free variable.

\subsection{Conversion and Equality in the De Bruijn Notation}

A pleasant property of the de Bruijn notation is that we get
$\alpha$-equivalence for free, \ie, any two terms which are
$\alpha$-equivalent in the lambda calculus have the same
representation in the de Bruijn notation. Thus the complicated
$\alpha$-equivalence check in the lambda calculus has been replaced by
a simple syntactic equality check in the de Bruijn notation.

We must also reconsider $\beta$-contraction in this notation. Given
the de Bruijn $\beta$-redex $((\lambdadb M)\app N)$ we want to think
about substituting $N$ for the first free variable in $M$. But in
performing this contraction, we have also eliminated a lambda which
was previously over the term $M$. Thus all the free variables in $M$
will have to have their index decremented by one. Also, we may have to
substitute $N$ beneath some lambdas which will require us to renumber
all the free variables in $N$. This leads us to consider a generalized
notion of substitution which allows us to substitute for every free
variable in a term.

\begin{defn}[De Bruijn substitution]
\label{def:debruijn-substitution}
Let $t$ be a de Bruijn term and let $s_1, s_2, s_3 \ldots$ be an
infinite sequence of de Bruijn terms. The result of simultaneously
substituting $s_i$ for the $i$th free variable in $t$ is denoted
by $S(t; s_1, s_2, s_3, \ldots)$ and is define by,
\begin{enumerate}
\item $S(c; s_1, s_2, s_3, \ldots) = c$, for any constant $c$,
\item $S(\#i; s_1, s_2, s_3, \ldots) = s_i$, for any index $\#i$,
\item $S((t_1\ t_2); s_1, s_2, s_3, \ldots) = S(t_1; s_1, s_2,
  s_3, \ldots)\ S(t_2; s_1, s_2, s_3, \ldots)$, and
\item $S((\lambdadb t); s_1, s_2, s_3, \ldots) = \lambdadb S(t; \#1,
  s_1', s_2', s_3', \ldots)$ where $s_i' = S(s_i; \#2, \#3, \#4,
  \ldots)$.
\end{enumerate}
\end{defn}

The interesting case in this definition is when we descend underneath
an abstraction. Within this abstraction, the index $\#1$ should be
left untouched, the index $\#2$ should refer to what we were
substituting for the first free variable, the index $\#3$ should refer
to what we were substituting for the second free variable,
etc. Further, since the arguments being substituted in are going to be
placed under this additional lambda, we need to increment the indices
of all free variables in those arguments. Given this extended
definition of substitution, we can define $\beta$-conversion for the
de Bruijn notation.

\begin{defn}[De Bruijn $\beta$-conversion]
\label{def:debruijn-beta}
A term $s$ results from a term $r$ by a $\beta$-contraction, denoted
$r \onebeta s$, if $s$ can be obtained by replacing a subterm of $r$
of the form $((\lambdadb t_1)\app t_2)$, referred to as a
$\beta$-redex, by $S(t_1; t_2, \#1, \#2, \ldots)$. We say also that
$s$ results from $r$ by a $\beta$-reduction, denoted $r \mbeta s$, if
it can be obtained from $r$ by a (possibly empty) sequence of
$\beta$-contractions.  The term $r$ is said to result from $s$ by a
$\beta$-expansion if $s$ results from $r$ by a $\beta$-contraction.
Finally, $r$ and $s$ are said to be $\beta$-equivalent of one results
from the other by a sequence of $\beta$-contractions and
$\beta$-expansions, and we denote this by $r =_\beta s$.
\end{defn}

While convenient from the implementation standpoint, the de Bruijn
notation is not particularly readable for humans. For example, the
term $(\lambda x . (\lambda y . (\lambda z . x)\ x)\ x)$ is encoded as
$(\lambdadb (\lambdadb (\lambdadb \#3)\ \#2)\ \#1)$. A naive glace at
this term may suggest that three different variables are referenced,
even though the same variable is referenced each time. Similarly, the
term $(\lambda x . (\lambda y . (\lambda z . z)\ y)\ x)$ is encoded as
$(\lambdadb (\lambdadb (\lambdadb \#1)\ \#1)\ \#1)$. Again, a naive
glace may suggest that the same variable is referenced three times
when in fact each reference is to a different variable. For the rest
of this thesis we shall use the de Bruijn notation, but because of the
reasons stated above we will often present examples using the named
lambda calculus.

\section{Properties of the Lambda Calculus}

The $\beta$-equivalence rule performs the ``heavy-lifting'' of the
lambda calculus. In the computational setting we use
$\beta$-equivalence to determine the value of a computation, and in
the representational setting we use $\beta$-equivalence to determine
if two terms represent the same thing. In this section we look at
properties of the lambda calculus which make it suitable for use in
both of these settings.

\subsection{\texorpdfstring{$\beta$}{Beta}-equivalence and
  Confluence}
\label{sec:confluence}

Determining $\beta$-equivalence is seemingly difficult because we can
use both $\beta$-contraction and $\beta$-expansion. Using
$\beta$-expansion is impractical since we can apply it anywhere in a
term, thus we will try to restrict ourselves to $\beta$-contraction
which limits us to considering only the $\beta$-redexes of a term. In
making this restriction, we may fear that we lose completeness, \ie,
that two terms are $\beta$-equivalent, but there is no common term to
which they $\beta$-reduce. The next theorem addresses this fear and
assures us that such a situation cannot occur.

Before we can state the theorem, we need to introduce a diagram
notation which is common in rewriting systems such as the lambda
calculus. In such systems, we often have statements of the form
``Let P and Q holds, then R and S are true'' where $P$, $Q$, $R$, and
$S$ denote some relationships between terms. We represent this in a
diagram by drawing solid arrows for the given properties, $P$
and $Q$, and using dashed arrows for the resulting properties, $R$ and
$S$.

\begin{theorem}[Church-Rosser property]
\label{thm:beta-church-rosser}
Let $t_1$ and $t_2$ be lambda terms such that $t_1 =_\beta t_2$. Then
there exists a term $t_3$ such that $t_1 \mbeta t_3$ and $t_2 \mbeta
t_3$, \ie, the follow diagram holds.
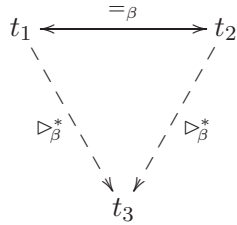
\begin{center}
\begin{minipage}{5cm}
\xymatrix{
t_1 \ar@{-->}[rdd]_{\mbeta} \ar@{<->}[rr]^{=_\beta} &
& t_2 \ar@{-->}[ldd]^{\mbeta} \\
\\
& t_3
}
\end{minipage}
\end{center}
\end{theorem}

This theorem tells us that there is no gap in completeness if we
restrict ourselves to $\beta$-reduction when determining
$\beta$-equivalence. This result is equivalent to the following
property,

\begin{theorem}[Confluence]
\label{thm:beta-confluent}
Let $t_1, t_2, t_3$ be lambda terms such that $t_1 \mbeta t_2$ and
$t_1 \mbeta t_3$. Then there exists a term $t_4$ such that $t_2
\mbeta t_4$ and $t_3 \mbeta t_4$, \ie, the following diagram
holds.
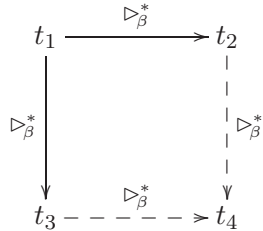
\begin{center}
\begin{minipage}{5cm}
\xymatrix{
t_1 \ar@{->}[rr]^{\mbeta} \ar@{->}[dd]_{\mbeta} &
& t_2 \ar@{-->}[dd]^{\mbeta} \\
\\
t_3 \ar@{-->}[rr]^{\mbeta} & & t_4
}
\end{minipage}
\end{center}
\end{theorem}

\subsection{Normal Forms and Typed Lambda Calculi}
\label{sec:typed}

The Church-Rosser property gives us some guidance in determining
$\beta$-equivalence by allowing us to consider only
$\beta$-reduction. A general method for determining
$\beta$-equivalence is still incomplete since the process of
$\beta$-reduction can go on indefinitely, such as for $((\lambda x .
x\app x)\app (\lambda x . x\app x))$. Thus we are often interested in
a subset of lambda terms for which $\beta$-contraction is no longer
applicable. This is the content of the following definition and
theorem.

\begin{defn}[$\beta$-normal form]
\label{def:beta-nf}
A lambda term is in $\beta$-normal form if it has no
$\beta$-redexes. If $t_1$ and $t_2$ are lambda terms such that
$t_1 \mbeta t_2$ and $t_2$ is in $\beta$-normal form, then we say that
$t_2$ is the $\beta$-normal form of $t_1$. When there is no ambiguity,
we may call this simply the normal form of $t_1$.
\end{defn}

\begin{theorem}
$\beta$-normal forms are unique.
\end{theorem}

An even stronger property than having a $\beta$-normal form is for a
term to be strongly $\beta$-normalizing. This means that any sequence
of $\beta$-reductions is terminating and therefore reaches the
$\beta$-normal form. When terms are strongly $\beta$-normalizing, we
can determine $\beta$-equivalence by reducing to $\beta$-normal form
and comparing for equality. Thus, when dealing with strongly
$\beta$-normalizing terms, we have a complete decision procedure.

One system for ensuring terms are strongly $\beta$-normalizing is the
simply typed lambda calculus \cite{church-40-types}. In this calculus
we have a set of types which we assign the constants in the
language. In addition, we annotate all lambdas with a type which
represents the valid argument type of the function. Using this, we can
build up and assign types to a whole range of terms. To begin with, we
define the types in our language.

\begin{defn}[Simple types]
Simple types are defined by
\[ T ::= \alpha\ |\ T \to T \]
where $\alpha$ ranges over a nonempty set of base types.
\end{defn}

We use the names $A$ and $B$ to represent types.  We assign a type $A$
to a term $t$ by creating a typing judgement which says that $A$ is a
valid type for $t$. The form of this judgement is $\Gamma \yields t :
A$ where $\Gamma$, called the context, contains type assignments for
the free variables and $\Sigma$, called the signature, contains type
assignments for the constants. The context maintains type assignments
for free variables using a stack of types
$A_1.A_2.\cdots.A_n.\emptyset$ which represent types for the free
variables $\#1$, $\#2$, \ldots, $\#n$, respectively. The rules for
constructing these typing judgements are given in
\autoref{fig:typed}. Notice that in these rules, the lambdas are
annotated with the type of their arguments.

The two key theorems about the simply typed lambda calculus are that
$\beta$-reduction preserves typing and that typed terms are strongly
$\beta$-normalizing.

\begin{theorem}[Preservation of types]
If $\Gamma \yields t : A$ and $t \onebeta t'$ then $\Gamma \yields t'
: A$.
\end{theorem}

\begin{theorem}[Strong normalization of typed terms]
If $\Gamma \yields t : A$ then $t$ is strongly $\beta$-normalizing.
\end{theorem}

\begin{figure}[t]
\begin{align*}
&\infer[\mbox{where $c : A \in \Sigma$}]{\Gamma \yields c : A}{}
&
&\infer{A.\Gamma \yields \#1 : A}{} \\
\\
&\infer[\mbox{where $i > 1$}]{A.\Gamma \yields \#i : A'}{\Gamma
  \yields \#(i-1) : A'}
&
&\infer{\Gamma \yields (t_1\app t_2) : A}{\Gamma \yields
  t_1 : B \to A & \Gamma \yields t_2 : B} \\
\\
&\infer{\Gamma \yields (\lambdadbann{A} t) : A \to B}{A.\Gamma
  \yields t : B}
\end{align*}
\caption{Typing rules for the simply typed lambda calculus}
\label{fig:typed} 
\end{figure}

The simply typed lambda calculus is only one example of a typing
system which may be layered on top of the lambda calculus.
Many other typing systems have been developed
and put to use in real systems.  For instance, $\lambda$Prolog
\cite{nadathur-91-overview} uses a polymorphic simply typed lambda
calculus \cite{nadathur-92-type} and Twelf \cite{pfenning-99-twelf}
uses the a dependently typed lambda calculus \cite{harper-93-LF}.

\section{Existential Variables and Substitution}
\label{sec:metavars}

The representational use of the lambda calculus produces a need for
variables which can be instantiated. For example, suppose we want to
express a rewrite rule in logic which says $\forall x . (F \wedge
G(x)) \rightarrow F \wedge \forall x .  G(x)$. Such a rule allows us
to pull a term out of a universal quantifier if it does not contain
the variable being quantified over. Using the ideas discussed in the
introduction of this thesis we can think of encoding the left-hand
side of this rule as $forall\ (\lambda x. F \wedge (G\app x))$ where
$F$ and $G$ are variables which will be instantiated upon matching
this rule with a specific instance. These variables are called {\it
  meta variables} and determining appropriate substitutions for them
is called higher-order unification.

The logical interpretation of meta variables requires that
substitutions cannot capture variables. Thus in our example above, if
$F$ or $G$ contain free occurrences of $x$, then the bound variable $x$
must be renamed before the substitutions are performed. This
restriction is actually a benefit to the use of meta variables because
it allows us to have precise control over the possible variable
occurrences within a substitution. For instance, although $G$ cannot
bind the variable $x$ in our example, $x$ is provided as an argument
so that $G$ can be of the form $(\lambda x. G')$ where $G'$ is a term
which has free occurrences of $x$. The process of $\beta$-contraction
then ties the knot and associates the free occurrences of $x$ in $G'$
with the argument $x$ in $(G\app x)$. For example, consider
substituting $(\lambda x.x + x > x)$ for $G$. Then the left-hand side
becomes
\[ forall\ (\lambda x. F \wedge ((\lambda x. x + x > x)\app x))
\onebeta forall\ (\lambda x. F \wedge (x + x > x)) \]
Conversely, since $x$ is not an argument to $F$ we know
that $F$ does not depend on $x$ and so moving the $F$ outside of the
quantifier $\forall x$ is a logically sound operation.


\chapter{The Suspension Calculus}
\label{ch:susp}

The lambda calculus revolves around $\beta$-contraction. In turn,
$\beta$-contraction depends on a monolithic substitution operation
which is impractical for use in actual implementations. A solution to
this is to move the substitution operation down from the meta level
into the object level, \ie, make it explicitly part of the syntax of
the lambda calculus. This allows for direct manipulation of
substitutions and therefore fine-grained control over the substitution
process. The focus of this chapter is on a specific explicit
substitution calculus: the suspension calculus.

The first half of this chapter, extending from
\autoref{sec:motivation} to \autoref{sec:susp-metavars}, introduces
the suspension calculus and possible variations on it.  We start this
introduction by motivating the notation used to encode substitutions
in the suspension calculus and defining notation and rules based on
this motivation. We then describe the relationship between the
suspension calculus of this thesis and the original suspension
calculus from which it is created.  Next we show how the suspension
calculus admits typing in the style of the simply-typed lambda
calculus. We then discuss different ways in which meta variables can
be added to the calculus.

The second half of this chapter, comprising the remaining sections,
deals with proving important properties of the suspension calculus.
First, we prove that the rules governing substitution are terminating,
thus reflecting the finite nature of substitution in the lambda
calculus. Second, we show that these substitution rules are confluent,
thus the choices we make in performing substitution always result in
the same normal form. Third, we show that the suspension calculus
faithfully models the process of $\beta$-reduction in the lambda
calculus. Fourth, we prove that the property of confluence extends to
the full suspension calculus, thus making it a candidate for new
approaches to unification. Finally, we prove a property intrinsic to
the suspension calculus which relates different ways of representing
substitutions.

\section{Motivation for the Encoding of Substitutions}
\label{sec:motivation}

Before we formally and explicitly define the syntax of the suspension
calculus, it is beneficial to consider what information needs to be
reflected into the syntax. Note that we are doing this reflection in
the context of the de Bruijn notation since it provides a unique
representation of $\alpha$-equivalent terms. The first difficultly in
this respect is that substitution in the de Bruijn notation is an
operation with infinitely many arguments. For instance we have the
$\beta$-contraction rule which say
\[ ((\lambdadb t_1)\app t_2) \onebeta S(t_1; t_2, \#1, \#2,
\ldots). \] The substitution here is for infinitely many variables and
thus no naive embedding of this into the syntax will work. Instead, it
helps to start with a fresh view of the substitution needed for
$\beta$-reduction.

Consider the following term for which we want to perform
$\beta$-reduction but delay its effect on $t$,
\begin{equation}
\label{eq:before}
(\ldots ((\lambdadb \ldots(\lambdadb \ldots ((\lambdadb \ldots t
\ldots)\ s_1)\ldots )\ldots)\ s_2) \ldots)
\end{equation}
Here we have a redex with $s_2$ as an argument and within the body of
this redex we have another redex with $s_1$ as an argument. We want to
consider the effect on $t$ of contracting these redexes. That is, we
wish to produce a term of the following form:
\begin{equation}
\label{eq:after}
(\ldots (\ldots(\lambdadb \ldots ( \ldots t' \ldots)\ldots )\ldots)
\ldots)
\end{equation}
where $t'$ is an encoding of the term $t$ together with the
information needed to perform the substitutions generated by
contracting the two redexes. We call $t'$ a {\it suspension} since it
represents a suspended substitution. The information in this
suspension consists of substitutions for some variables and
renumberings for the other variables. In order to express this
information, it helps to distinguish between two types of variables in
$t$: those that referred to a variable bound within the outermost
lambda that is contracted and those that are free with respect to
the outermost lambda. We use the outermost lambda as the reference
point for this information since we hope to make all information
local to the contractions being made, \ie, the context of our redexes
should not affect the term $t'$. In the discussion that follows, we
shall fixate on the term starting with the outermost lambda that is
contracted and ignore the context in which it occurs.

For the variables that are free there is a renumbering which must be
done to account for the lambdas which occurred before the reduction
that are gone after the reduction. In order to define this, we
introduction the idea of an {\it embedding level} which is the depth
of a term counted by the number of lambdas between it and the
top level. The embedding level of $t$ in (\ref{eq:before}) is called the
{\it old embedding level}, $ol$, and the embedding level of $t'$ in
(\ref{eq:after}) is called the {\it new embedding level}, $nl$. For
example, if the lambdas shown are the only lambdas in the term, then
old embedding level is 3 and the new embedding level is 1. The number
of lambdas that are removed is $ol - nl$ and thus this is the
renumbering to be done on the free variables.  Furthermore, we can
easily find the free variables since they are the ones whose index is
greater than their embedding level, \eg, if $\#3$ occurs at embedding
level 2 then we know it occurs underneath two lambdas and represents
the first free variable outside of these lambdas.

In addition to renumberings for the free variables, we must provide
substitutions for the bound variables. In the case that a bound
variable does not need a substitution (since its binding lambda is not
contracted) we will create a dummy substitution which substitutes the
first free variable, $\#1$, thus preserving the term.  Now, the terms
within substitutions often come from a different embedding level than
where we are thinking of substituting them and must be renumbered so
that their free variables are not captured by the context into which
we substitute. In order to do this, for each substitution we keep a
number indicating the embedding level from which it came, say $l$.
Then when we need to perform a substitution we increment all free
variables in the substituted term by $nl - l$. We keep these
substitutions together with their embedding levels in a list, ordered
from the first bound variable to the last. This allows us to simply
add a dummy substitution to the front of this list in order to shift
all indices when we descend underneath an abstraction.

Given the previous information for encoding substitutions, we write
our encoding as $\env{t,ol,nl,e}$ where $t$ is the term over which
substitution is performed, $ol$ is the old embedding level, $nl$ is
the new embedding level, and $e$, called the environment, is a list of
substitutions for the bound variables. The overall term
$\env{t,ol,nl,e}$ is called a suspension. If we looked further into
$e$, its structure would be $(t_1,l_1)::(t_2,l_2)::\ldots::nil$ where
the $t_i$ are terms and $l_i$ are the corresponding embedding levels.

The general operation of the suspension calculus will be to perform
$\beta$-contractions which produce suspensions and then to apply
rewriting rules which move these suspensions deeper and deeper into
the tree until they are applied to a final term. During this, it is
natural that we might encounter a term of the form $\env{\env{t, ol_1,
    nl_1, e_1}, ol_2, nl_2, e_2}$, which represents a term $t$
together with two substitutions to be performed on it sequentially. We
can think of merging these two substitutions to produce a term of the
form $\env{t, ol', nl', e'}$, \ie, a single suspension which encodes
the information of both of the previous ones. This requires careful
consideration of the values for $ol'$ and $nl'$ together with new
syntax to represent the shape of $e'$.

We can determine values for $ol'$ and $nl'$ by thinking carefully
about the embedding levels in the term $\env{\env{t, ol_1, nl_1, e_1},
  ol_2, nl_2, e_2}$. Here we have two substitutions over $t$ which
possibly overlap with each other. In detail, the effect on $t$ is
first to substitute for the first $ol_1$ free variables and then to
raise all the free variables by $nl_1$. This raising has the effect of
embedding the term within $nl_1$ abstractions.  Then the second
substitution walks over the resulting term, substituting for the first
$ol_2$ free variables and then raising all the free variables by
$nl_2$. Note that some of these $ol_2$ substitutions will be vacuous
since the free variables have all been raised by $nl_1$. In fact, if
$nl_1 \geq ol_2$, then all of the substitutions of the second
suspension are vacuous. In this case we have $ol' = ol_1$ since only
the first $ol_1$ substitutions are performed. Also, $nl' = nl_2 +
(nl_1 - ol_2)$ since the raising done is that of $nl_2$ and $nl_1$
minus the $ol_2$ vacuous substitutions. In the case when $nl_1 <
ol_2$, we have $ol' = ol_1 + (ol_2 - nl_1)$ since we have the $ol_1$
substitutions and all but the first $nl_1$ of the $ol_2$
substitutions. Also, $nl' = nl_2$ since all of the $nl_1$ raisings are
consumed by the $ol_2$ substitutions and so the only raising left over
is from the second suspension. These two branching cases for
$ol'$ and $nl'$ can be coalesced by using the minus operator on
natural numbers. Then we have $ol' = ol_1 + (\monus{ol_2}{nl_1})$ and
$nl' = nl_2 + (\monus{nl_2}{ol_1})$ in both cases.

We must also determine the shape of $e'$ after merging
$\env{\env{t,ol_1,nl_1,e_1}, ol_2, nl_2, e_2}$ into
$\env{t,ol',nl',e'}$. The result should roughly be the substitutions
of $e_1$, modified by the substitutions in $e_2$, together with some
tail portion of $e_2$. For each term in $e_1$, we can use $nl_1$ to
compute the number of abstractions in which it is
embedded. Using this we can prune off the first elements of $e_2$
which correspond to these abstractions. When we have done this
for all elements of $e_1$, we only have left to determine which tail
portion of $e_2$ to include. The length of this tail portion should be
the number of abstractions consumed by the second suspension, $ol_2$,
minus the number of abstractions created by the first
suspension, $nl_1$. Thus we can compute the total shape of $e'$ by
knowing only $e_1$, $e_2$, $ol_2$, and $nl_1$. We write the resulting
form as $\menv{e_1,nl_1,ol_2,e_2}$ and we call this a {\it merged
  environment}.

\section{Syntax of the Suspension Calculus}

In this section we formally define the syntax of the suspension
calculus and present measures for assessing the wellformedness of
expressions in this calculus. We begin by defining the ``pre-syntax''
of suspension expressions, which is the raw syntax without any
constraints on wellformedness.

\begin{defn}[Pre-syntax of suspension expressions]
\label{def:presyntax}
The ``pre-syntax'' of suspension expressions is given by the following
definitions of the syntactic categories of ``terms'' and
``environments.''
\begin{tabbing}
\qquad\=$t$\=\quad::=\quad\=\kill
\>$t$\>\quad::=\quad\>$c \ \vert \ \#i\ \vert\ (t\app t)\ \vert\
(\lambdadb t)\ \vert\ \env{t, n, n, e}$ \\
\>$e$\> \quad::=\quad \> $nil\ \vert\ ((t,l) :: e)\ \vert\ \menv{e, n,
  n, e}$
\end{tabbing}
Here $c$ ranges over an enumerable set of constants, $i$ ranges over
the natural numbers, and $n$ and $l$ ranges over the non-negative
integers.
\end{defn}

We call $\#i$ a variable index or reference, and we call $(t\app t)$
and $(\lambdadb t)$ application and abstraction, respectively. The
term $\env{t,n,n,e}$ is called a {\it suspension}. The operation $::$
is a consing operator on lists and the $(t,l)$ component of $((t,l) ::
e)$ is called an {\it environment term}. Finally, $\menv{e,n,n,e}$ is
called a {\it merged environment}. We collectively refer to all terms,
environments, and environment terms generated by the above definition
as {\it suspension expressions}. We will drop parenthesis by assuming
application is left associative, the scope of a lambda extends as far
right as possible, and $::$ is right associative. We will also assume
the obvious definition of subexpression.

In order to move from pre-syntax to syntax, we need to consider
constraints on expressions so that they ``make sense.'' For example,
in a term of the form $\env{t,ol_1,nl_1,e_1}$ we are thinking of
performing substitution for the first $ol_1$ free variables using the
substitutions in $e_1$. Thus the number of substitutions in $e_1$,
called the length of $e_1$, must be $ol_1$. Also, for each
substitution in $e_1$ we include an embedding level from which the
substitution came. Since substitutions will always move inward, it
must be that we keep moving to deeper embedding levels. Thus our
current embedding level, $nl_1$, must be greater than or equal to the
embedding level of every environment term in $e_1$. We enforce this by
defining a measure called the level of $e_1$. These measures are the
content of the following definitions.

\begin{defn}[Length of an environment]
\label{def:len}
The length of an environment $e$ is denoted by $len(e)$ and is defined
recursively by
\begin{align*}
len(nil) &= 0 \\
len((t,n) :: e) &= 1 + len(e) \\
len(\menv{e_1, nl_1, ol_2, e_2}) &= len(e_1) + (\monus{ol_2}{nl_1})
\end{align*}
\end{defn}

\begin{defn}[Level of an environment or environment term]
\label{def:lev}
The level of an environment $e$ is denote $lev(e)$ and is defined
recursively by
\begin{align*}
lev(nil) &= 0 \\
lev((t,n) :: e) &= n \\
lev(\menv{e_1, nl_1, ol_2, e_2}) &= lev(e_2) + (\monus{nl_1}{ol_2})
\end{align*}
\end{defn}

Given these two definitions, we can define the syntax of suspension
expressions
\begin{defn}[Syntax of suspension expressions]
\label{def:syntax}
The syntax of suspension expressions is those expressions in
\autoref{def:presyntax} with the following additional wellformedness
constraints,
\begin{enumerate}
\item In any subexpression of the form $\env{t,ol,nl,e}$, we
  must have $len(e) = ol$ and $lev(e) \leq nl$.
\item In any subexpression of the form $\menv{e_1,nl_1,ol_2,e_2}$, we
  must have $len(e_2) = ol_2$ and $lev(e_1) \leq nl_1$.
\item In any subexpression of the form $(t,n)::e$, we must have
  $lev(e) \leq n$.
\end{enumerate}
\end{defn}

\section{Rules of the Suspension Calculus}

Now that we have created syntax which embeds substitutions we can
consider rules which operate on this modified syntax. The effect of
these rules should be in three parts: (1) to contract beta redexes to
produce suspensions, (2) to move these suspensions down in the syntax
tree until they can be applied, and (3) to merge suspensions and
compute the resulting merged environment. The complete rules of the
suspension calculus are listed in \autoref{fig:susprules} and are
divided into three categories described above: the $\beta_s$ rules,
the reading rules, and the merging rules.

The $\beta_s$ rule simulates $\beta$-contraction in the suspension
calculus using the suspension syntax to encode the effect of
substitution. This rule rewrites the $\beta$-redex $((\lambdadb
t_1)\app t_2)$ to $\env{t_1, 1, 0, (t_2, 0)::nil}$ which says that we
substitute $t_2$ in for the first free variable in $t_1$ and decrement
all other free variables by one.

The second category of rules is the reading rules, (r1)-(r6), which
provide a means for moving suspensions down in a term and also
performing substitutions. Taken together, the $\beta_s$ rule and the
reading rules form an adequate simulation of the de Bruijn calculus, as
we shall prove in \autoref{sec:simulation}.

The merging rules, (m1)-(m6), are the final category of rules. The
(m1) rule enables us to merge two suspension into a single suspension,
at the cost of creating a merged environment. The rules (m2)-(m6)
then allow us to evaluate this merged environment in a lazy way.

\begin{figure}[t]
\begin{tabbing}
\quad\=(r11)\quad\=\kill
\> ($\beta_s$)\>$((\lambdadb t_1)\app t_2) \ra \env{t_1, 1, 0, (t_2,0)
  :: nil}$.\\[7pt] 
\> (r1)\> $\env{c,ol,nl,e} \ra c$, provided $c$ is a constant.\\[5pt]
\> (r2)\>$\env{\#i,0,nl,nil} \ra \#j$, where $j = i + nl$.\\[5pt] 
\> (r3)\> $\env{\#1,ol,nl,(t,l) :: e} \ra \env{t,0,nl',nil}$, where
$nl' = nl - l$.\\[5pt] 
\> (r4)\> $\env{\#i, ol, nl, (t,l) :: e} \ra \env{\# i',
  ol',nl,e},$\\
\>\> where $i' = i - 1$ and $ol' = ol -1$, provided $i > 1$.\\[5pt] 
\> (r5)\> $\env{(t_1\app t_2),ol,nl,e} \ra (\env{t_1,ol,nl,e}\app
\env{t_2,ol,nl,e})$.\\[5pt] 
\> (r6)\>$\env{(\lambdadb t), ol, nl, e} \ra (\lambdadb \env{t,
  ol', nl', (\#1, nl') :: e})$,\\
\>\> where $ol' = ol + 1$ and $nl' = nl + 1$.\\[7pt]
\> (m1)\>$\env{\env{t,ol_1,nl_1,e_1},ol_2,nl_2,e_2} \ra
\env{t,ol',nl', \menv{e_1,nl_1,ol_2,e_2}}$,\\
\>\>where $ol' = ol_1 + (\monus{ol_2}{nl_1})$ and $nl' = nl_2 +
(\monus{nl_1}{ol_2})$.\\[5pt]
\> (m2)\>$\menv{e_1, nl_1, 0, nil} \ra e_1$.\\[5pt] 
\> (m3)\>$\menv{nil, 0, ol_2, e_2} \ra e_2$.\\[5pt] 
\> (m4)\>$ \menv{nil, nl_1, ol_2, (t,l) :: e_2} \ra \menv{nil, nl_1',
  ol_2', e_2}$,\\
\>\> where $nl_1' = nl_1 - 1$ and $ol_2' = ol_2 - 1$, provided $nl_1
\geq 1$.\\[5pt]
\> (m5)\> $ \menv{(t,n) :: e_1, nl_1, ol_2, (s,l) ::
  e_2} \ra \menv{(t,n) :: e_1, nl_1', ol_2', e_2}$,\\
\>\>where $nl_1' = nl_1 - 1$ and $ol_2' = ol_2 - 1$, provided $nl_1 >
n$.\\[5pt] 
\> (m6)\> $\menv{(t,n) :: e_1, n, ol_2, (s,l) :: e_2} \ra
(\env{t, ol_2, l, (s,l) :: e_2}, m) :: \menv{e_1, n, ol_2, (s,l) :: e_2}$,\\
\>\>where $m = l + (\monus{n}{ol_2})$.
\end{tabbing}
\caption{Rewrite rules for the suspension calculus}
\label{fig:susprules}
\end{figure}

\begin{defn}
The reduction relations generated by the rules in
\autoref{fig:susprules} are denoted by $\onebetas$, $\oneread$, and
$\onemerge$. The relations $\onereadmerge$, $\onereadbetas$, and
$\oneall$ are the appropriate unions of those relations. If $R$
corresponds to any of these relations then we will use $R^*$ to denote
its reflexive and transitive closure.
\end{defn}

The following example illustrates a use of these rules where $t$,
$s_1$, and $s_2$ are arbitrary suspension expressions. This example is
a simplified version of (\ref{eq:before}).
\begin{align*}
& (\lambdadb \lambdadb ((\lambdadb t)\app s_1))\app s_2 \\
&\betas \env{\lambdadb \env{t, 1, 0, (s_1,0)::nil}, 1, 0, (s_2,
  0)::nil} \\
&\oneread \lambdadb \env{\env{t, 1, 0, (s_1, 0)::nil}, 2, 1, (\#1,
  1)::(s_2, 0)::nil} \\
&\onemerge \lambdadb \env{t, 3, 1, \menv{(s_1, 0)::nil, 0, 2, (\#1,
    1)::(s_2, 0)::nil}} \\
&\merge \lambdadb \env{t, 3, 1, (\env{s_1, 2, 1, (\#1,
    1)::(s_2,0)::nil}, 1)::(\#1,1)::(s_2,0)::nil}
\end{align*}
The outermost suspension here encodes three substitutions to be made
over $t$. The first substitution is $s_1$, modified by substituting
$s_2$ for its second free variable. The second substitution is a dummy
substitution which corresponds to the lambda that remains after
contraction. Finally, the last substitution corresponds to
substituting in $s_2$ and since the embedding level of this is one
less than the new embedding level, we will have to raise all the free
variables of $s_2$ which corresponds to our substituting of $s_2$
underneath a lambda.

In order for our rules to make sense, they need to produce terms
that make sense. In formal terms, we need to ensure that
using a rule on a well-formed expression produces a well-formed
expression.

\begin{theorem}
Let $e$ be a well-formed suspension expression and let $e \oneall
e'$. Then $e'$ is a well-formed suspension expression.
\end{theorem}
\begin{proof}
This property must be proved simultaneously with two other properties:
if $e$ is an environment then $lev(e) \geq lev(e')$ and $len(e) =
len(e')$. The reason is that if we could rewrite an environment so
that its level increases or so its length changes, then we might break
the wellformedness of an expression such as $\env{t,ol,nl,e}$ where
$lev(e) \leq nl$ and $len(e) = ol$ by rewriting $e$ to $e'$ such that
$lev(e') > nl$ or $len(e) \neq ol$. The proof of all three properties
simultaneously is a simple case analysis on the rewrite from $e$ to
$e'$. We give two examples here in order to give the flavor of the
argument.

Consider (r3) and suppose that the left hand side is well-formed. This
means that $ol = len((t,l)::e)$ and $nl \geq lev((t,l)::e) = l$. In
order for the right hand side to be well-formed we must have $0 =
len(nil)$ and $nl' \geq lev(nil) = 0$. The first is trivially true,
and the second requires that we show $nl - l \geq 0$ which follows
from $nl \geq l$.

As a second example, consider (m6). Assuming the left hand side is
well formed yields $n \geq lev((t,n)::e_1)$, $ol_2 = len((s,l)::e_2)$,
and $lev(et) \geq lev(e)$. In order to show the right hand side is
well-formed we must have $l \geq lev((s,l)::e_2)$, $ol_2 =
len((s,l)::e_2)$, $lev(et) \geq lev((s,l)::e_2)$, $m \geq
lev(\menv{e_1,n,ol_2,(s,l)::e_2}$, $n \geq lev(e_1)$, $ol_2 =
len((s,l)::e_2)$, and $l \geq lev(e_2)$. All of these follow
directly. Since (m6) is a rule on environments, we must also verify
that this rewriting does not cause the level of this environment to
increase. The level of the left hand side is $lev((s,l)::e_2) +
(\monus{n}{ol_2})$ and the level of the right hand side is $m = l +
(\monus{n}{ol_2})$. Since $l = lev((s,l)::e_2)$, this follows
easily. Finally, it is plain to see that the length is preserved.
\end{proof}

From here on we will speak only of well-formed suspension expressions
and thus drop the ``well-formed'' label on them. The final definition
and lemma of this section are ones that allows us to massage
environments into a convenient form for use with the rewrite
rules. This will be of importance in \autoref{sec:original} and
\autoref{sec:confluence-rm}.

\begin{defn}[Simple environments and truncating]
A simple environment is one of the form $(t_0,l_0) :: (t_1,l_1) ::
\ldots :: (t_{n-1}, l_{n-1}) :: nil$, with $n$ possibly zero.  For $0
\leq i < n$, we write $e\{i\}$ to denote the truncated environment
with the first $i$ elements removed, \ie, $(t_i,l_i) :: \ldots ::
(t_{n-1},l_{n-1}) :: nil$, with $i$ possibly zero. We extend this
notation by letting $e\{i\}$ denote $nil$ in the case that $i \geq
len(e)$ for any simple environment $e$.
\end{defn}

\begin{lemma}
\label{lem:eval-simple}
Let $e$ be an environment. Then there exists a simple environment $e'$
such that $e \merge e'$.
\end{lemma}
\begin{proof}
The proof is by case analysis on the structure of $e$, basically
showing that if $e$ is not a simple environment then we can always
apply a rule (m2)-(m6) to it. Connecting this to the result requires
that we also know that the rewrite rules are terminating, which is
shown in \autoref{sec:termination}.
\end{proof}

\section{Relationship to the Original Suspension Calculus}
\label{sec:original}

The current suspension calculus is based on the original suspension
calculus by Nadathur and Wilson \cite{nadathur-98-suspension}. There
are two differences between these calculi: the way dummy substitutions
are handled in the rule (r6) and the way merging is performed using
(m2)-(m6). In this section we highlight these differences and explain
the connection between these two calculi. Before we begin, however, we
note that the original suspension calculus had a separate syntactic
class for environments terms because it had more possibilities for
such expressions. Hence, in this section we will treat environment
terms as a separate syntactic class in both calculi.

The first difference is how dummy substitutions are handled in the
case of rule (r6). When pushing a suspension underneath an abstraction
there is a need to generate a substitution for the bound
variable. Since we are not actually contracting a redex, this
substitution should have no net effect and is thus a
dummy substitution compared to those substitutions generated by
$\beta$-contraction. In the original calculus, the rule for pushing a
suspension underneath an abstraction had the form
\[ \env{\lambdadb t, ol, nl, e} \ra \lambdadb \env{t, ol+1, nl+1,
  @nl::e} \] This new $@nl$ environment term was conceived of as an
optimization to separate real and dummy substitutions, and it also has
the benefit of simplifying the proof of termination for the reading
and merging rules. Nevertheless, we can simplify the calculus
significantly by using the environment term $(\#1, nl+1)$
instead. This environment term has the same effect the same effect and
also allows us to have a simpler system since we can exclude all the
old rules for manipulating $@nl$ forms. With this change, the results
of the original suspension calculus paper still hold, which we will
assume \cite{nadathur-98-suspension}.

The larger difference between the two calculi is the way merging is
performed using (m2)-(m6). Taking into account the change above, the
$\beta_s$ and reading rules are the same in both calculi, but the
merging rules are still significantly different. The merging rules for
the original calculus are presented in \autoref{fig:original}. The
rule (m8') makes use of a measure called the index which we will say
more about later in this section. For now, it is sufficient to think
of the index as a measure very similar to the level. Pay special
attention to (m5') which says
\[ \menv{et::e_1, nl_1, ol_2, e_2} \ra \menvt{et, nl_1, ol_2, e_2} ::
\menv{e_1, nl_1, ol_2, e_2} \] The effect of this rule is to eagerly
propagate the effects of $e_2$ onto the environment term $et$ by
creating a new environment term $\menvt{et, nl_1, ol_2, e_2}$. This
new form is then used to prune $e_2$ using rules (m6') and (m7') until
only the portion relevant to $et$ was left. In the current calculus,
this pruning is done using the form $\menv{e_1, nl_1, ol_2, e_2}$
using rule (m5) and thus the work of this pruning is shared for each
environment term of $e_1$. This change in the current calculus allows
us to use fewer syntactic forms and also fewer rules, both of which
reduce the energy required to understand the calculus.

A possible downside of these simplifications is that we might lose
some desirable theoretical properties of the calculus. This turns out
not to be the case as we prove later in this chapter. In fact, the
simplifications to the calculus allow new results such as a typed
version of the calculus in \autoref{sec:typed-suspension} and a
translation to another explicit substitution calculus in
\autoref{sec:lambdasigma}. In the rest of this section, we will look
at a more formal relationships between our calculus and its
predecessor.

\begin{figure}[t]
\begin{tabbing}
\qquad (r11)\qquad\=\kill \qquad (m1')\>
$\env{\env{t,ol_1,nl_1,e_1},ol_2,nl_2,e_2} \ra
\env{t,ol',nl', \menv{e_1,nl_1,ol_2,e_2}}$,\\
\>where $ol' = ol_1 + (\monus{ol_2}{nl_1})$ and $nl' = nl_2 +
(\monus{nl_1}{ol_2})$.\\
[5pt]
\qquad (m2')\>$\menv{nil, nl, 0, nil} \ra nil$.\\
[5pt]
\qquad (m3')\> $\menv{nil, 0, ol, e} \ra e$.\\
[5pt] \qquad (m4')\>$\menv{nil, nl, ol, et :: e} \ra \menv{nil,
  nl', ol', e}$,\\
\>where $nl,ol \geq 1$, $nl' = nl - 1$ and $ol' = ol - 1$.\\
[5pt] \qquad (m5')\> $\menv{et :: e_1, nl,ol,e_2} \ra \menvt{et,nl,ol,e_2}
:: \menv{e_1,nl,ol,e_2}$.\\
[5pt]
\qquad (m6')\> $\menvt{et,nl,0,nil} \ra et$.\\
[5pt] \qquad (m7') \>$\menvt{et,nl,ol,et' :: e} \ra \menvt{et, nl',ol', e},$\\
\>where $nl' = nl - 1$ and $ol' = ol - 1$, provided $ind(et) < nl$.\\
[5pt] \qquad (m8')\>$\menvt{(t,nl),nl,ol,et :: e} \ra (\env{t,ol,l',et :: e}, m)$\\
\>where $l' = ind(et)$ and $m = l' + (\monus{nl}{ol})$.
\end{tabbing}
\caption{Merging rules for the original suspension calculus}
\label{fig:original}
\end{figure}

Suspension expressions in our context are a subset of the original
suspension expressions. The only difficultly in showing this is that
wellformedness rules for the original suspension calculus are defined
the same as in \autoref{def:syntax} except with the measure
level replaced by the measure index, defined as follows.

\begin{defn}[Index of an environment or environment term]
\label{def:ind}
Given a natural number $i$, the $i$-th index of an environment $e$ is
denoted by $ind_{i}(e)$ and is defined as follows:
\begin{enumerate}
\item If $e$ is $nil$ then $ind_{i}(e) = 0$.
\item If $e$ is $(t,k) :: e'$
then $ind_{i}(e)$ is $k$ if $i = 0$ and $ind_{i-1}(e')$ otherwise.
\item If $e$ is $\menv {e_1,nl,ol,e_2 }$, let
 $m = (\monus{nl}{ind_{i}(e_1)})$ and $l = len(e_1)$. Then
\begin{tabbing}
\qquad\=\kill
\>$ind_{i}(e) = \left\{ \begin{array}{ll}
      ind_{m}(e_2) + (\monus{nl}{ol}) &
                \mbox{ if $i < l$ and  $len(e_2) > m$}\\ 
      ind_{i}(e_1)   & \mbox{ if $i < l$ and  $len(e_2) \leq  m$}\\
      ind_{(i - l+nl)}(e_2) & \mbox{ if $i \geq l$.}
                        \end{array} 
               \right.$
\end{tabbing}
\end{enumerate}
The index of an environment, denoted by $ind(e)$, is
$ind_{0}(e)$.
\end{defn}

Intuitively, the index of an environment is the embedding level of its
first environment term and zero if the environment is nil. The index
of an environment term $et$ is the value of $n$ for which $et \merge
(t,n)$.  We can consider the index measure on expressions in the
current calculus as well. Note that this measure is equal to the level
on environments terms in the current calculus because they already
have the form $(t,n)$. The only difference in the current calculus
between index and level is in the case of merged environments, where
the level is an upper bound on the index.

\begin{lemma}
Let $e$ be a well-formed environment in the current calculus. Then
$lev(e) \geq ind(e)$.
\end{lemma}
\begin{proof}
First generalize to $lev(e) \geq ind_i(e)$ for all $i$. Then the proof
proceeds by induction on the structure of $e$.
\end{proof}

\begin{lemma}
Well-formed terms in the current suspension calculus are well-formed
in the original suspension calculus.
\end{lemma}
\begin{proof}
This is a direct consequence of the previous lemma. For example, if
$\env{t, ol, nl, e}$ is a well-formed term of the current calculus
then we have $ol = len(e)$ and $nl \geq lev(e) \geq ind(e)$. Thus it
is a well-formed term in the original suspension calculus.
\end{proof}

We now know that the well-formed expressions we are working with in
the current calculus are also well-formed in the original calculus. We
can also show that the rules of the current calculus are either
derived or admissible rules for the original calculus. We will
consider each rule of the current calculus in turn, ignoring rules
that are the same in both calculi, \ie, (m1), (m3), and (m4).

The rule (m2) operates on an environment of the form $\menv{e_1, nl_1,
  0, nil}$. By \autoref{lem:eval-simple}, the environment $e_1$ can
be rewritten to the form $et_0::et_1::\ldots::et_{n-1}::nil$.  Then by
applying (m5') $n$ times we can get
\[ \menvt{et_0, nl_1, 0, nil} :: \menvt{et_1, nl_1, 0, nil}
:: \ldots :: \menvt{et_{n-1}, nl_1, 0, nil} :: \menv{nil, nl_1, 0,
  nil} \]
Using (m2') and $n$ applications of (m6') this rewrites to $et_0 ::
et_1 :: \ldots :: et_{n-1} :: nil$. Thus both $\menv{e_1, nl_1, 0, nil}$
and $e_1$ can rewrite to a common term and therefore the rule (m2) is
admissible.

For rule (m5) we cite the original suspension paper where Lemma 6.10
states that $\menv{e_1, nl + 1, ol + 1, et::e_2}$ and $\menv{e_1, nl,
  ol, e_2}$ rewrite to a common term if $ind(e_1) \leq nl$
\cite{nadathur-98-suspension}. Restricting this to the case where
$e_1 = (t,n)::e_1'$ and restating the requirement as $n < nl + 1$
yields the rule (m5). Thus (m5) is admissible.

Finally consider (m6). Assuming $\menv{(t,n)::e_1, n, ol_2, et::e_2}$
is a term in the current suspension calculus, we can rewrite it using
(m5') followed by (m8') to produce
\[ (\env{t,ol_2,l,et::e_2},m)::\menv{e_1, n, ol_2, et::e_2} \] where
$l = ind(et)$ and $m = l + (\monus{n}{ol_2})$. Since $et$ is an
environment term in the current calculus we have $l = lev(et)$ and
thus (m6) is a derived rule of the original calculus.

We can formalize our observations into the following theorem which
will be useful in \autoref{sec:simulation} when we show that the
current calculus properly simulates $\beta$-contraction.

\begin{theorem}[Same normal form]
\label{thm:same-nf}
Let $x$ be a well-formed (current) suspension expression. Then the
$\onereadmerge$-normal form of $x$ is also the $\onereadmergep$-normal
form of $x$.
\end{theorem}
\begin{proof}
This theorem depends on result that the reading and merging rules are
confluent and terminating in both calculi (See
\autoref{sec:termination} and \autoref{sec:confluence-rm} for the
current calculus, \cite{nadathur-98-suspension} for the original
calculus). The result then follows by induction on the rewrite
sequence which takes $x$ to its $\onereadmerge$-normal form.
\end{proof}

\section{Typed Version of the Suspension Calculus}
\label{sec:typed-suspension}

In this section we present a typed version of the suspension calculus
which was not previously possible with the original suspension
calculus. This typed version is is consistent with the simply-typed
lambda calculus from \autoref{sec:typed} and motivated in the same
way. In fact our typing judgment for terms of the
form $c$, $(\lambdadb t)$, and $(t_1\app t_2)$ is the same as in the
simply-typed lambda calculus, but significant complexity is introduce
to handle the new judgment for $\env{t,ol,nl,e}$. The issue is that
we must interpret the term $t$ in the context of the substitutions
encoded in $e$ and relative to $nl$. This necessitates a new judgment
which talks about the effect of an environment (relative to an
embedding level) on a typing context. The form of this judgment is
$\Gamma \yields e \one{nl} \Gamma'$. Where $\Gamma$ and $\Gamma'$ are
contexts, $\Sigma$ is a signature, $e$ is an environment, and $nl$ is
an integer for the embedding level. All of typing rules are presented
in \autoref{fig:typed-suspension}.

\begin{figure}[t]
\begin{align*}
&\infer[\mbox{where $c : A \in \Sigma$}]{\Gamma \yields c : A}{}
&
&\infer{A.\Gamma \yields \#1 : A}{} \\
\\
&\infer[\mbox{where $i > 1$}]{A.\Gamma \yields \#i : A'}{\Gamma
  \yields \#(i-1) : A'}
&
&\infer{\Gamma \yields (t_1\app t_2) : A}{\Gamma \yields
  t_1 : B \to A & \Gamma \yields t_2 : B} \\
\\
&\infer{\Gamma \yields (\lambdadbann{A} t) : A \to B}{A.\Gamma
  \yields t : B}
&
&\infer{\Gamma \yields \env{t,ol,nl,e} : A}{\Gamma \yields
  e \one{nl} \Gamma' & \Gamma' \yields t : A} \\
\\ \\
&\infer{\Gamma \yields nil \one{0} \Gamma}{}
&
&\infer[\mbox{where $nl > 0$}]{A.\Gamma \yields nil \one{nl}
  \Gamma'}{\Gamma \yields nil \one{nl-1} \Gamma'} \\
\\
&\infer[\mbox{where $nl > n$}]{A.\Gamma \yields (t,n)::e \one{nl}
  \Gamma'}{\Gamma \yields (t,n)::e \one{nl-1} \Gamma'}
&
&\infer{\Gamma \yields (t,n)::e \one{n} A.\Gamma'}{\Gamma \yields t :
  A & \Gamma \yields e \one{n} \Gamma'} \\
\\
&\infer{\Gamma \yields \menv{e_1, nl', ol', e_2} \one{nl}
  \Gamma''}{\Gamma \yields e_2 \one{nl - (\monus{nl'}{ol'})}
  \Gamma' & \Gamma' \yields e_1 \one{nl'} \Gamma''} \\
\end{align*}
\caption{Typing rules for the typed suspension calculus}
\label{fig:typed-suspension} 
\end{figure}

The following result establish the wellformedness of our typing
judgments.

\begin{theorem}
Type judgments are preserved by the rewrite relations
\end{theorem}
\begin{proof}
The critical thing to show is that if $\ell \to r$ is a rewrite rule
and $\Gamma \yields \ell : A$ (respectively $\Gamma \yields \ell
\one{nl} \Gamma'$) then $\Gamma \yields r : A$ (respectively $\Gamma
\yields r \one{nl} \Gamma'$). The proof proceeds by cases on the
rewrite rule and we focus here on a few interesting cases.

Let the rewrite rule be $(\beta_s)$. Then we are given that $\Gamma
\yields ((\lambdadbann{A} t_1)\app t_2) : B$ from which we know the
derivation tree must be
\[
\infer{\Gamma \yields ((\lambdadbann{A} t_1)\app t_2) : B}{
  \infer{\Gamma \yields (\lambdadbann{A} t_1) : A \to B}{
    \infer{A.\Gamma \yields t_1 : B}{D_1}}
  && \infer{\Gamma \yields t_2 : A}{D_2}}
\]
Where $D_1$ and $D_2$ are appropriate derivations. We can then use
these derivations to construct the following typing judgment.
\[
\infer{\Gamma \yields \env{t_1,1,0,(t_2,0)::nil}}{
  \infer{\Gamma \yields (t_2,0)::nil \one{0} A.\Gamma}{
    \infer{\Gamma \yields t_2 : A}{D_2}
    && \Gamma \yields nil \one{0} \Gamma}
  && \infer{A.\Gamma \yields t_1 : B}{D_1}}
\]
This completes the argument for $(\beta_s)$.

Consider the case of (r3). Then we must have the following typing
derivation.
\[
\infer{A_{nl}.A_{nl-1}.\cdots.A_{l+1}.\Gamma \yields
  \env{\#1,ol,nl,(t,l)::e} : A}{
  \infer{A_{nl}.A_{nl-1}.\cdots.A_{l+1}.\Gamma \yields
    (t,l)::e \one{nl} A.\Gamma'}{
    \infer{A_{nl-1}.\cdots.A_{l+1}.\Gamma \yields
      (t,l)::e \one{nl-1} A.\Gamma'}{
      \infer{\vdots}{
        \infer{\Gamma \yields (t,l)::e \one{l} A.\Gamma'}{
          \infer{\Gamma \yields t : A}{D_1}
          && \infer{\Gamma \yields e \one{l} \Gamma'}{D_2}}}}}
  && A.\Gamma' \yields \#1 : A}
\]
From this we can construct the typing judgment,
\[
\infer{A_{nl}.A_{nl-1}.\cdots.A_{l+1}.\Gamma \yields
  \env{t,0,nl-l,nil} : A}{
  \infer{A_{nl}.A_{nl-1}.\cdots.A_{l+1}.\Gamma \yields nil \one{nl-l}
    \Gamma}{
    \infer{A_{nl-1}.\cdots.A_{l+1}.\Gamma \yields nil \one{nl-l-1}
      \Gamma}{
      \infer{\vdots}{\Gamma \yields nil \one{0} \Gamma}}}
  && \infer{\Gamma \yields t : A}{D_1}}
\]

As a final example, consider the rule (m3) which yields the following
typing derivation.
\[
\infer{\Gamma \yields \menv{nil, 0, ol_2, e_2} \one{nl} \Gamma'}{
  \infer{\Gamma \yields e_2 \one{nl} \Gamma'}{D_1}
  && \Gamma' \yields nil \one{0} \Gamma'}
\]
And this directly contains the needed typing judgment.
\[
\infer{\Gamma \yields e_2 \one{nl} \Gamma'}{D_1}
\]
\end{proof}

\section{Meta Variables and the Suspension Calculus}
\label{sec:susp-metavars}

The syntax of suspension expressions does not presently allow for meta
variables, as described in \autoref{sec:metavars}. We can remedy
this by adding modifying the syntax of terms to
\begin{tabbing}
\qquad\=$t$\=\quad::=\quad\=\kill
\>$t$\>\quad::=\quad\>$v\ \vert\ c \ \vert \ \#i\ \vert\ (t\app t)\ \vert\
(\lambdadb t)\ \vert\ \env{t, n, n, e}$,
\end{tabbing}
where $v$ represents the category of meta variables. Because such
variables have a logical interpretation, any substitution for them
must avoid capturing local variables. This means that any outer
context cannot effect the value of such variables. To accommodate this
interpretation, we can add the following to our reading rules:
\begin{tabbing}
\qquad (r7)\qquad\=\kill
\qquad (r7)\> $\env{v,ol,nl,e} \ra v$, if $v$ is a meta variable.
\end{tabbing}
This rule has the same character as the (r1) rule for constants, and
as a result, all the properties of the suspension calculus without
logical meta variables carry over to the suspension calculus with
logical meta variables.

An alternative interpretation for meta variables in one in which
substitution is performed without renaming to avoid variable
capture. This is referred to as the {\it graftable} interpretation of
meta variables, and it has been found useful in higher-order
unification procedures \cite{dowek-95-higherorder}. The benefit of
using graftable meta variables is that dependencies of meta variables
on the outside context no longer need to be explicitly specified,
which turns out to be a significant cost in the traditional
higher-order unification algorithm \cite{huet-75-unification}. On the
other hand, it may seem that using graftable meta variables removes
any possibility of control over dependencies, but we can in fact
enforce some restrictions using explicit raising. For example, to
prevent the metavariable $X$ from depending the the de Bruijn indices
$\#1$ and $\#2$, we can replace it with the term
$\env{X,0,2,nil}$. This also us to simulate logical meta variables
using graftable meta variables simply by lifting such graftable
variables so that they can not depend on any of the bound variables in
the term.

In order to support graftable meta variables in the calculus, we
extend the syntax as before, but instead of adding the rule (r7), we
leave the rules unchanged. This is consistent with the graftable
interpretation: because we know nothing about such meta variables, we
cannot say what effect a suspension will have on them. Adding
graftable meta variables to the calculus, however, introduces some new
complications with respect to confluence. For example, consider the
term $((\lambdadb ((\lambdadb X)\app t_1))\app t_2)$ in which $X$ is a
graftable meta variable and $t_1$ and $t_2$ are terms in
$\onereadmerge$-normal form. This term can be rewritten to
\begin{tabbing}
\qquad\=\kill
\> $\env{\env{X,1,0,(t_1,0)::nil},1,0,(t_2,0)::nil}$
\end{tabbing}
and also to 
\begin{tabbing}
\qquad\=\kill
\> $\env{\env{X,2,1,(\#1,1)::(t_2,0)::nil},1,0,
      (\env{t_1,1,0,(t_2,0)::nil},0)::nil}$,
\end{tabbing}
amongst other terms. It is easy to see that these terms cannot now be
rewritten to a common form using only the reading and $(\beta_s)$
rules. The merging rules are essential to this ability and, as we see
in \autoref{sec:confluence-rm}, these also suffice for this
purpose. Another impact of graftable meta variables is that normal
forms with respect to the reading and merging rules now include the
possibility of remaining suspensions, whereas without graftable meta
variables such normal forms are always de Bruijn terms.

We assume henceforth that the suspension calculus includes meta
variables under the graftable interpretation. For reasons already
mentioned, it is easy to see that the properties we establish for the
resulting calculus will hold also under the logical interpretation.

\section{Termination of Reading and Merging Rules}
\label{sec:termination}

The first significant property we prove for the suspension calculus is
that the reading and merging rules are terminating, \ie, that all
$\onereadmerge$-sequences are finite. This is useful in all future
sections because it allows us to induct on $\onereadmerge$-sequences,
and it tells that $\onereadmerge$-normal forms always exist. On a
deeper level, the substitution process in the lambda calculus is
inherently finite and by showing that our elaboration of this
substitution process is also finite, we argue convincingly for the
well behaved nature of our rules.

We prove the termination of the reading and merging rules in two
steps. First, we describe a collection of first-order terms and a
wellfounded order on those terms using a lexicographic recursive path
ordering \cite{dershowitz-82-orderings,
  ferreira-94-wellfoundedness}. Second, we define a mapping from
suspension expressions to newly described terms such that each of the
reading and merging rules produces a smaller term with respect to the
defined order. The desired conclusion follows from these facts. The
key points of this work have been verified in Coq.\footnote{\url{http://www-users.cs.umn.edu/~agacek/pubs/gacek-masters/Termination/}}

We imagine the terms we describe here to be an abstract view of the
suspension calculus such that only details relevant to the termination
of the reading and merging rules are considered. These terms 
are constructed using the following (infinite) vocabulary:
the following (infinite) vocabulary: the 0-ary function symbol {\it
  *}, the unary function symbol {\it lam}, and the binary function
symbols {\it app}, {\it cons} and, for each positive number $i$,
$s_i$. We denote this collection of terms by $\cal T$. We assume the
following partial ordering $\sqsupset$ on the signature underlying
$\cal T$: $s_i \sqsupset s_j$ if $i > j$ and, for every $i$, $s_i
\sqsupset {\it app}$, $s_i \sqsupset {\it lam}$, $s_i \sqsupset {\it
  cons}$ and $s_i \sqsupset {\it *}$. This ordering is now extended to
the collection of terms.

\begin{defn}[Term Order]
\label{def:termorder}
The relation $\succ$ on $\cal T$ is inductively defined by the
following property: Let $s = f(s_1,\ldots,s_m)$ and $t =
g(t_1,\ldots,t_n)$; both $s$ and $t$ may be {\it *}, \ie, the number
of arguments for either term may be $0$. Then $s \succ t$ if 
\begin{enumerate}
\item $f = g$ (in which case $n = m$), $(s_1, \ldots, s_n) \succ_{lex}
  (t_1,\ldots,t_n)$, and, $s \succ t_i$ for all $i$ such that $1 \leq
  i \leq n$, or
\item $f \sqsupset g$ and $s \succ t_i$ for all $i$ such that $1 \leq
  i \leq n$, or
\item $s_i = t$ or $s_i \succ t$ for some $i$ such that $1 \leq i \leq
  m$.
\end{enumerate}
Here $\succ_{lex}$ denotes the lexicographic ordering induced by
$\succ$. 
\end{defn}

In the terminology of \cite{ferreira-94-wellfoundedness}, $\succ$ is
an instance of a recursive path ordering based on $\sqsupset$. It is
easily seen that $\sqsupset$ is a well-founded ordering on the
signature underlying $\cal T$. The results in
\cite{ferreira-94-wellfoundedness} then imply the following:

\begin{lemma}
\label{lem:succ-wellfounded}
$\succ$ is a well-founded partial order on $\cal T$.
\end{lemma}

We now consider the translation from suspension expressions to $\cal
T$. The critical part of this mapping is the treatment of expressions
of the form $\env{t,ol,nl,e}$ and $\menv{e_1,nl,ol,e_2}$. Because of
(m1) and (m6), there is a tight relationship between the encoding of
these two types of expressions. It turns out that we can ignore the
differences between them when looking at our abstract view of the
suspension calculus, thus we drop the $ol$ and $nl$ from each and
encode them as $s_i$ for some $i$.

To determine the appropriate value for $i$ in $s_i$, we must consider
how this $i$ will be needed. We will focus on the case for
$\env{t,ol,nl,e}$, but the same ideas will carry over to
$\menv{e_1,nl,ol,e_2}$. A first attempt to translate $\env{t,ol,nl,e}$
as $s_i$ for some fixed $i$ would fail since the rule (r3) would not
yield a smaller term when applied due to the lexicographic component
of our ordering. Instead, we use the the value of $i$ as a coarse
measure of the remaining substitution work, so that this value
decreases in the rule (r3). In order for this to work we must count
the $\#1$ on the left-hand side of (r3) for some positive amount. This
then passes the problem onto (r6) where we add a $\#1$ to the
right-hand side. In order to balance this, we assign lambdas a weight
based on the number of suspensions in which they are embedded. This
results in our family of measures $\eta_j$ where $j$ represents the
number of levels of suspensions or merged environments that the
current term is embedded underneath. In defining this measure we must
be aware that the rule (m1) allows the environment $e_2$ to become
embedded underneath an additional level. Thus the embedding level of
an environment must be based on the number of suspensions in the term
over which the environment applies. We call this count of suspensions
the ``internal embedding potential'' and denote it by $\mu$ in the
following definitions. In these definitions, {\it max} is the function
that picks the larger of its two integer arguments.

\begin{defn}\label{def:intembedding} 
The measure $\mu$ that estimates the
  internal embedding potential of a suspension expression is defined as
  follows:
\begin{enumerate}
\item For a term $t$, $\mu(t)$ is $0$ if $t$ is a constant, a meta
  variable or a de Bruijn index, $\mu(s)$ if $t$ is $(\lambdadb
  s)$, ${\it  max}(\mu(s_1),\mu(s_2))$ if $t$ is $(s_1\app
  s_2)$, and  $\mu(s) + \mu(e) + 1$ if $t$ is
  $\env{s,ol,nl,e}$.  

\item For an environment $e$, $\mu(e)$ is $0$ if $e$ is nil,
  ${\it max}(\mu(s),\mu(e_1))$ if $e$ is $(s,l) :: e_1$ and
  $\mu(e_1) + \mu(e_2) + 1$ if $e$ is $\menv{e_1,nl,ol,e_2}$.
\end{enumerate}
\end{defn}

\begin{defn}\label{def:suspsize}
The measures $\eta_i$ on terms and environments for each natural
number $i$ are defined simultaneously by recursion as follows:
\begin{enumerate}
\item For a term $t$, $\eta_i(t)$ is $1$ if $t$ is a constant, a meta
  variable or a de Bruijn index, $\eta_i(s) + 1$ if $t$ is $(\lambdadb
  s)$, ${\it  max}(\eta_i(s_1),\eta_i(s_2)) + 1$ if $t$ is $(s_1\app
  s_2)$,and  $\eta_{i+1}(s) + \eta_{i+1+\mu(s)}(e) + 1$ if $t$ is
  $\env{s,ol,nl,e}$.  

\item For an environment $e$, $\eta_i(e)$ is $0$ if $e$ is nil,
  ${\it max}(\eta_i(s),\eta_i(e_1))$ if $e$ is $(s,l) :: e_1$ and
  $\eta_{i+1}(e_1) + \eta_{i+1+\mu(e_1)}(e_2) + 1$ if $e$ is
  $\menv{e_1,nl,ol,e_2}$.
\end{enumerate}
\end{defn}

\begin{defn}\label{def:susptoessence}
The translation $\ess$ of suspension expressions to $\cal
T$ is defined as follows:
\begin{enumerate}
\item For a term $t$, $\ess(t)$ is {\it *} if $t$ is a
  constant a meta variable or a de Bruijn index,
  ${\it app}(\ess(t_1),\ess(t_2))$ if $t$ is $(t_1\app t_2)$,
  ${\it lam}(\ess(t'))$ if $t$ is $(\lambdadb t')$ and
  $s_i(\ess(t'),\ess(e'))$ where $i = \eta_0(t)$ if $t$ is
  $\env{t',ol,nl,e'}$. 

\item For an environment $e$, $\ess(e)$ is {\it *} if $e$ is
  {\it nil}, ${\it cons}(\ess(t'),\ess(e'))$ if $e$ is
  $(t',l) :: e'$ and $s_i(\ess(e_1), \ess(e_2))$ where $i =
  \eta_0(e)$ if $e$ is $\menv{e_1,nl,ol,e_2}$.
\end{enumerate}
\end{defn}

Using this translation we now lift our ordering on this collection of
first-order terms to suspension expressions.

\begin{defn}[Suspension Expression Order]
\label{def:exprorder}
For suspension expressions $s$ and $t$ we say $s \gg t$ if and only if
$\ess(s) \succ \ess(t)$.
\end{defn}

There key properties of the $\succ$ relation carry over to the $\gg$
relation. First, subexpressions are smaller than their parent
expressions. Second, the relation is monotonic in the sense that if
$v$ results from $u$ by replacement of a subpart $x$ by $y$ such that
$x \gg y$, then $u \gg v$. Third, the relation is wellfounded.  These
properties together with the following theorem will make this relation
a powerful tool for performing induction over suspension expressions.

\begin{theorem}
\label{thm:rm-termination}
Every rewriting sequence based on the reading and merging rules
terminates.
\end{theorem}
\begin{proof}
A tedious but straightforward inspection of each of the reading and
merging rules verifies the following: If $l \ra r$ is an instance of
these rules, then $l \gg r$, $\mu(l) \geq \mu(r)$, and, for every
natural number $i$, $\eta_i(l) \geq \eta_i(r)$.  Further, it is easily
seen that if $x$ and $y$ are both either terms or environments such
that $\mu(x) \geq \mu(y)$ and $\eta_i(x) \geq \eta_i(y)$ for each
natural number $i$ and if $v$ is obtained from $u$ by substituting $y$
for $x$, then $\eta_i(u) \geq \eta_i(v)$ for each natural number
$i$. From these observations it follows easily that if $t_1
\onereadmerge t_2$ then $t_1 \gg t_2$. The theorem is
now a consequence of \autoref{lem:succ-wellfounded}.
\end{proof}

\section{Confluence of Reading and Merging Rules}
\label{sec:confluence-rm}

In this section we prove that the reading and merging rules are
confluent, \ie, that the choices we make in rewriting can always be
reconciled. Thus $\onereadmerge$-normal forms are unique, which is
another argument for the coherence of our reading and merging rules.

The property of confluence states that given terms $f$, $g$, and $h$
such that $f \readmerge g$ and $f \readmerge h$, there exists a term
$k$ such that $g \readmerge k$ and $h \readmerge k$. This can be
expressed using the diagrams described in \autoref{sec:confluence}
as,
\begin{center}
\begin{minipage}{5cm}
\xymatrix{
f \ar@{->}[dd]_{\readmerge} \ar@{->}[rr]^{\readmerge} &&
g \ar@{-->}[dd]^{\readmerge} \\
\\
h \ar@{-->}[rr]^{\readmerge} && k \\
}
\end{minipage}
\end{center}
A well known result (proved, for instance, in
\cite{huet-80-confluent}) in rewriting is that a terminating rewriting
system is confluent if it is weakly confluent. Weak confluence is
states that given terms $f$, $g$, and $h$ such that $f \onereadmerge
g$ and $f \onereadmerge h$, there exists a term $k$ such that $g
\readmerge k$ and $h \readmerge k$, \ie, that the following figure
holds.
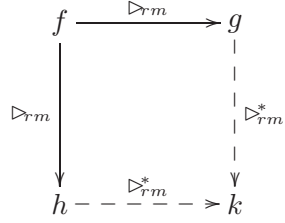
\begin{center}
\begin{minipage}{5cm}
\xymatrix{
f \ar@{->}[dd]_{\onereadmerge} \ar@{->}[rr]^{\onereadmerge} &&
g \ar@{-->}[dd]^{\readmerge} \\
\\
h \ar@{-->}[rr]^{\readmerge} && k \\
}
\end{minipage}
\end{center}
This is much easier to show since we only need to consider one rewrite
step from $f$ to $g$ and from $f$ to $h$. In doing this, we must
consider each possible overlap between any two rules. The most
complicated of these is the overlap of (m1) with itself when applied
to a term of the form
\[ \env{\env{\env{t, ol_1, nl_1, e_1}, ol_2, nl_2, e_2}, ol_3, nl_3,
e} \] This leads us to develop an associativity property for
merged environments which is the content of the following
section.

\subsection{An Associativity Property for Environment Merging}
\label{sec:assoc}

Here we show that the following two environments rewrite to a
common environment.
\[ A = \menv{\menv{e_1, nl_1, ol_2, e_2}, nl_2 +
  (\monus{nl_1}{ol_2}), ol_3, e_3} \]
\[ B = \menv{e_1, nl_1, ol_2 + (\monus{ol_3}{nl_2}), \menv{e_2,
    nl_2, ol_3, e_3}} \]
Essentially this tells us that to compute the effect of $e_2$ on
$e_1$ and then compute the effect of $e_3$ on the result is the
same as computing the effect of $e_3$ on $e_2$ and then computing
the effect of that result on $e_1$. Ignoring the details for a moment,
suppose that $e_1 = (t_1,n_1)::e_1'$ and we are able to apply the rule
(m6) to both terms (that is twice to $A$ and once to $B$). Then the
term portion of the environment term for $A$ is roughly 
\[ \env{\env{t_1,ol_2,nl_2,e_2},ol_3,nl_3,e_3} \]
and for $B$ it is roughly
\[ \env{t_1, ol_2 + (\monus{ol_3}{nl_2}), nl_3 + (\monus{nl_2}{ol_3}),
  \menv{e_2,nl_2,ol_3,e_3}} \]
Then we can apply (m1) to bring these two terms back together. This is
the heart of the proof. The vast majority of the proof is taken up by
details and corner cases. For instance, we might not be able to apply
(m6) to a term $\env{(t_1,n_1)::e_1,nl_1,ol_2,e_2}$ because $nl_1 >
n_1$, $e_2$ is nil, or $e_2$ isn't of the form $(t_2,n_2)::e_2'$. All
of these cases must be handled and often doubly so since we have to
apply (m6) twice to the term $A$. In order to ease these pains, we
introduce a few lemmas which we can use to massage expressions into the
proper form.

The rules (m4), (m5), and (m6) require their second environment to be
of the form $(t,n)::e$ which isn't the case with merged environments like
in the term $B$. Ideally we could use the rules (m5) and (m6) to turn
a term of the form $\menv{e,nl,ol,e'}$ into one of the form $(t,n)::e$,
but if we do this we will not be able to use our inductive hypothesis.
To accommodate this issue we introduce the following lemmas which
essentially state that applying the rules (m5) and (m6) does not
interfere with the process of rewriting.

\begin{lemma}\label{lem:prune-back} 
Let $A$ be the environment $\menv{e_1, nl_1, ol_1, \menv{e_2, nl_2,
ol_3, e_3}}$ where $e_3$ is a simple environment and $e_2$ is of
the form $(t_2,n_2) :: e'_2$. Further, for any positive number $i$ such
that $i \leq nl_2 - n_2$ and $i \leq ol_3$, let $B$ be the environment
\begin{tabbing}
\qquad\=\kill
\>$\menv{e_1, nl_1, ol_1, \menv{e_2, nl_2 - i, ol_3 - i,
    e_3\{i\}}}$.
\end{tabbing}
If $A \readmerge C$ for any simple environment $C$ then also $B
\readmerge C$.
\end{lemma}

\begin{proof} It suffices to verify the claim when $i = 1$; an easy
  induction on $i$ then extends the result to the cases where $i >
  1$. For the case of $i = 1$, the argument is by induction on the
  length of the reduction sequence from $A$ to $C$ with the
  essential part being a consideration of the first 
  rule used. The details are straightforward and hence
  omitted. 
\end{proof}

\begin{lemma}\label{lem:unfold-back} 
Let $A$ be the environment $\menv{e_1, nl_1, ol_1, \menv{e_2, nl_2,
ol_3, e_3}}$ where $e_2$ and $e_3$ are environments of the form
$(t_2,nl_2) :: e'_2$ and $(t_3,n_3) :: e'_3$, respectively. Further,
let $B$ be the environment 
\begin{tabbing}
\qquad\=\kill
\>$\menv{e_1, nl_1, ol_1, (\env{t_2,ol_3,n_3,e_3}, n_3 +
  (\monus{nl_2}{ol_3}))::\menv{e'_2,nl_2, ol_3, e_3}}$.
\end{tabbing}
If $A \readmerge C$ for any simple environment $C$ then also $B
\readmerge C$.
\end{lemma}

\begin{proof}
The proof is again by induction on the length of the reduction
sequence from $A$ to $C$. The first rule in this sequence either
produces $B$, in which case the lemma follows immediately, or it can
be used on $B$ (perhaps at more than one place) to produce a form that
is amenable to the application of the induction hypothesis. 
\end{proof}

In evaluating the composition of $e_2$ and $e_3$, it may be the case
that some part of $e_3$ is inconsequential. The last observation that
we need is that this part can be ``pruned'' immediately in calculating
the composition of the combination of $e_1$ and $e_2$ with $e_3$. The
following lemma is consequential in establishing this fact. 

\begin{lemma}\label{lem:remvac}
Let $A$ be the environment $\menv{e_1,nl_1,ol_2,e_2}$ where
$e_2$ is a simple environment.
\begin{enumerate}
\item If $ol_2 \leq nl_1 - {\it lev}(e_1)$ then $A$ reduces to any
  simple environment that $e_1$ reduces to.
\item For any positive number $i$ such that $i \leq nl_1 - {\it
  lev}(e_1)$ and $i \leq ol_2$, $A$ reduces to any simple environment
  that $\menv{e_1,nl_1 - i, ol_2 - i, e_2\{i\}}$ reduces to.
\end{enumerate}
\end{lemma}

\begin{proof} Let $e_1$ be reducible to the simple environment
  $e'_1$. Then we may transform $A$ to the form
  $\menv{e'_1,nl_1,ol_2,e_2}$. Recalling that the level of an
  environment is never increased by rewriting, we have that ${\it
  lev}(e'_1) \leq {\it lev}(e_1)$. From this it follows that
  $A$ can be rewritten to $e'_1$ using rules (m5) and (m2) if $ol_2
  \leq nl_1 - {\it lev}(e_1)$. This establishes the first part of the
  lemma.  

The second part is nontrivial only if $nl_1 - {\it lev}(e_1)$ and
  $ol_2$ are both nonzero. Suppose this to be the case and let $B$ be
  $\menv{e_1,nl_1-1,ol_2-1,e_2\{1\}}$. The desired result follows by
  an induction on $i$ if we can show that $A$ can be rewritten to any
  simple environment that $B$ reduces to. We do this by an induction
  on the length of the reduction sequence from $B$ to the simple
  environment. This sequence must evidently be of length at least
  one. If a proper subpart of $B$ is rewritten by the first rule in
  this sequence, then the same rule can be applied to $A$ as well and
  the induction hypothesis easily yields the desired conclusion. If
  $B$ is rewritten by one of the rules (m3)-(m6), then it must be the
  case that $A \onereadmerge B$ via either rule (m4) or (m5) from
  which the claim follows immediately. Finally, if $B$ is
  rewritten using rule (m2), then $ol_2 \leq nl_1 - {\it
  lev}(e_1)$. The second part of the lemma is now a consequence of the
  first part.
\end{proof}

We now prove the associativity property for environment
composition: 

\begin{lemma}
\label{lem:assoc}
Let $A$ and $B$ be environments of the form
\begin{tabbing}
\qquad\=\kill
\>$\menv{\menv{e_1, nl_1, ol_2, e_2}, nl_2 + (\monus{nl_1}{ol_2}),
  ol_3, e_3} $
\end{tabbing}
and
\begin{tabbing}
\qquad\=\kill
\>$\menv{e_1, nl_1, ol_2 + (\monus{ol_3}{nl_2}), \menv{e_2, nl_2,
    ol_3, e_3}}$,
\end{tabbing}
respectively. Then there is a simple environment $C$ such that $A
\readmerge C$ and $B \readmerge C$.
\end{lemma}

\begin{proof}
We assume that $e_1$, $e_2$ and $e_3$ are simple environments; if this
is not the case at the outset, then we may rewrite them to such
a form in both $A$ and $B$ before commencing the proof we provide. Our
argument is now based on an induction on the structure of $e_3$ with
possibly further inductions on the structures of $e_2$ and $e_1$. 

\medskip

\noindent {\it Base case for first induction.} When $e_3$ is {\it
  nil}, the lemma is seen to be true by observing that both $A$ and
  $B$ rewrite to $\menv{e_1,nl_1,ol_2,e_2}$ by virtue of rule (m2). 

\medskip

\noindent {\it Inductive step for first induction.} Let $e_3 =
(t_3,n_3) :: e'_3$. We now proceed by an induction on the structure
of $e_2$. 

\smallskip

\noindent {\it Base case for second induction.} When $e_2$ is {\it
  nil}, it can be seen that, by virtue of rules (m2), (m3) and either
  (m4) or (m5), $A$ and $B$ reduce to $\menv{e_1,nl_1,ol_3   - nl_2,
  e_3\{nl_2\}}$ when $ol_3 > nl_2$ and to $e_1$ otherwise. The
  truth of the lemma follows immediately from this.

\smallskip

\noindent {\it Inductive step for second induction.} Let $e_2 = (t_2,n_2) ::
e'_2$. We consider first the situation where $nl_1 >
lev(e_1)$. Suppose further that $ol_3 \leq (nl_2 - n_2)$. Using rules
(m5) and (m2), we see then that 
\begin{tabbing}
\qquad\=\kill
\>$B \readmerge \menv{e_1,nl_1,ol_2,e_2}$.
\end{tabbing}
We also note that $ol_3 \leq (nl_2 + (\monus{nl_1}{ol_2})) -
lev(\menv{e_1,nl_1,ol_2,e_2})$ in this case. \autoref{lem:remvac}
assures us now that $A$ can be rewritten to any simple
environment that $\menv{e_1,nl_1,ol_2,e_2}$ reduces to and thereby
verifies the lemma in this case.

It is possible, of course, that $ol_3 > (nl_2 - n_2)$. Here we see
that   
\begin{tabbing}
\qquad\=\kill
\>$B \readmerge \menv{e_1,nl_1 - 1, ol_2 + (\monus{ol_3}{nl_2}) - 1,
  \menv{e'_2,n_2,ol_3 - (nl_2 - n_2),e_3\{nl_2 - n_2\}}}$.
\end{tabbing}
using rules (m5) and (m6). Using rule (m5), we also have that
\begin{tabbing}
\qquad\=\kill
\>$A \readmerge \menv{\menv{e_1,nl_1 - 1, ol_2 - 1, e'_2},nl_2 +
  (\monus{nl_1}{ol_2}),ol_3,e_3}$.
\end{tabbing}
Invoking the induction hypothesis, it follows that $A$ and 
\begin{tabbing}
\qquad\=\kill
\>$\menv{e_1,nl_1 - 1, ol_2 + (\monus{ol_3}{nl_2}) -
  1,\menv{e'_2,nl_2,ol_3,e_3}}$ 
\end{tabbing}
reduce to a common simple environment. By \autoref{lem:prune-back}
it follows that $B$ must also reduce to this environment.

The only remaining situation to consider, then, is that when $nl_1 = 
lev(e_1)$. For this case we need the last induction, that on the
structure of $e_1$.

\smallskip

\noindent {\it Base case for final induction.} If $e_1$ is {\it nil},
  then $nl_1$ must be $0$. It follows easily that both $A$ and $B$
  reduce to $\menv{e_2,nl_2,ol_3,e_3}$ and that the lemma must
  therefore be true.

\smallskip

\noindent {\it Inductive step for final induction.} Here $e_1$ must be of the
form $(t_1,nl_1) :: e'_1$. We dispense first with the
situation where $n_2 < nl_2$. In this case, by rule (m5)
\begin{tabbing}
\qquad\=\kill
\>$B \readmerge
\menv{e_1,nl_1,ol_2+(\monus{ol_3}{nl_2}),\menv{e_2,nl_2 - 1, ol_3 - 1,
    e'_3}}$.
\end{tabbing}
By the induction hypothesis used relative to $e'_3$, $B$ and the
expression 
\begin{tabbing}
\qquad\=\kill
\>$\menv{\menv{e_1,nl_1,ol_2,e_2},nl_2 + (\monus{nl_1}{ol_2}) - 1,ol_3
  - 1, e'_3}$
\end{tabbing}
must reduce to a common simple environment. By \autoref{lem:remvac},
$A$ must also reduce to this environment. 

Thus, it only remains for us to consider the situation in which $n_2 =
nl_2$. In this case by using rule (m1) twice we may transform $A$ to
the expression $A_h :: A_t$ where 
\begin{tabbing}
\qquad\=\kill
\>$A_h = (\env{\env{t_1,ol_2,n_2,e_2},ol_3,n_3,e_3}, n_3 + (\monus{(nl_2 +
  (\monus{nl_1}{ol_2}))}{ol_3}))$
\end{tabbing}
and
\begin{tabbing}
\qquad\=\kill
\>$A_t = \menv{\menv{e_1', nl_1, ol_2, e_2}, nl_2 +
  (\monus{nl_1}{ol_2}), ol_3, e_3}$.
\end{tabbing}
Similarly, $B$ may be rewritten to the expression $B_h :: B_t$ where  
\begin{tabbing}
\qquad\=$B_h = $\=$($\=\qquad\qquad\=\kill
\>$B_h =$\>$($\>$\lenv t_1, ol_2 + (\monus{ol_3}{nl_2}), n_3 +
(\monus{nl_2}{ol_3}),$ \\
\>\>\>\>$(\env{t_2, ol_3, n_3, e_3}, n_3 +
  (\monus{nl_2}{ol_3}))::\menv{e_2',nl_2,ol_3,e_3} \renv,$\\
\>\>\>$n_3 + (\monus{nl_2}{ol_3}) + (\monus{nl_1}{(ol_2 +
  (\monus{ol_3}{nl_2}))}))$
\end{tabbing}
and
\begin{tabbing}
\qquad\=\kill
\> $B_t = \menv{e_1', nl_1, ol_2 + (\monus{ol_3}{nl_2}),
  (\env{t_2,ol_3,n_3,e_3}, n_3 + (\monus{nl_2}{ol_3}))::\menv{e_2',
    nl_2, ol_3, e_3}}$.
\end{tabbing}
Now, using straightforward arithmetic identities, it can be seen that
the ``index'' components of $A_h$ and $B_h$ are equal. Further, the
term component of $A_h$ can be rewritten to a form identical to the
term component of $B_h$ by using the rules (m1) and (m6). Finally, by
virtue of the induction hypothesis, it follows that $A_t$ and the
expression 
\begin{tabbing}
\qquad\=\kill
\>$\menv{e_1', nl_1, ol_2 +
(\monus{ol_3}{nl_2}), \menv{e_2, nl_2, ol_3, e_3}}$
\end{tabbing}
reduce to a common simple environment. \autoref{lem:unfold-back}
allows us to conclude that $B_t$ can also be rewritten to this
expression. Putting all these observations together it is seen that
$A$ and $B$ can be reduced to a common simple environment in this case
as well.
\end{proof}

\subsection{Proof of Confluence for Reading and Merging Rules}

We are now in a position to prove the confluence of the reading and
merging rules, using the ideas outlined in the beginning of this
section.

\begin{lemma}
\label{lem:rm-weak-confluence}
The relation $\onereadmerge$ is weakly confluent.
\end{lemma}
\begin{proof}
We recall the method of proof from \cite{huet-80-confluent}. An
expression $t$ constitutes a nontrivial overlap of the rules $R_1$ and
$R_2$ at a subexpression $s$ if (a)~$t$ is an instance of the lefthand
side of $R_1$, (b)~$s$ is an instance of the lefthand side of $R_2$
and also does not occur within the instantiation of a variable on the
lefthand side of $R_1$ when this is matched with $t$ and (c)~either
$s$ is distinct from $t$ or $R_1$ is distinct from $R_2$. Let $r_1$ be
the expression that results from rewriting $t$ using $R_1$ and let
$r_2$ result from $t$ by rewriting $s$ using $R_2$. Then the pair
$\langle r_1, r_2 \rangle$ is called the conflict pair corresponding
to the overlap in question. Relative to these notions, the theorem can
be proved by establishing the following simpler property: for every
conflict pair corresponding to the reading and merging rules, it is
the case that the two terms can be rewritten to a common form using
these rules.

In completing this line of argument, the nontrivial overlaps that we
have to consider are those between (m1) and each of the rules
(r1)-(r6), between (m1) and itself and between (m2) and (m3).  The
last of these cases is easily dealt with: the two expressions
constituting the conflict pair are identical, both being $nil$. The
overlap between (m1) and itself occurs over a term of the form
$\env{\env{\env{t,ol_1,nl_1,e_1},ol_2,nl_2,e_2},ol_3,nl_3,e_3}$. By
using rule (m1) once more on each of the terms in the conflict pair,
these can be rewritten to expressions of the form $\env{t,ol',nl',e'}$
and $\env{t,ol'',nl'',e''}$, respectively, whence we can see that $ol'
= ol''$ and $nl'=nl''$ by simple arithmetic reasoning and that $e'$
and $e''$ reduce to a common form using \autoref{lem:assoc}. The
overlaps between (m1) and the reading rules are also easily dealt
with. For instance, consider the case of (m1) and (r2) where we
have \[ \env{\env{\#i,0,nl_1,nil},ol_2,nl_2,e_2} \] Rewriting with
(r2) first produces \[ \env{\#(i+nl_1),ol_2,nl_2,e_2} \] while
rewriting with (m1) first yields \[ \env{\#i, \monus{ol_2}{nl_1},
  nl_2+(\monus{nl_1}{ol_2}), \menv{nil,nl_1,ol_2,e_2}}. \] For both
expressions we can rewrite $e_2$ to a simple environment $e_2'$ using
\autoref{lem:eval-simple}. Now if $ol_2 \geq nl_1$ then both terms
can be reconciled to \[ \env{\#i,ol_2 - nl_1, nl_2, e_2'\{nl_1\}} \]
In the case of $ol_2 < nl_1$, both terms rewrite to
$\#(i+nl_1-ol_2)$. Thus the conflict pair is resolved.  The other
cases of overlaps between (m1) and the reading rules are similar and
require roughly the same reasoning.
\end{proof}

As observed already, the main result of this section follows directly
from \autoref{lem:rm-weak-confluence} and \autoref{thm:rm-termination}.

\begin{theorem}
\label{th:rm-confluence}
The relation $\onereadmerge$ is confluent.
\end{theorem}

The uniqueness of $\onereadmerge$-normal forms is an immediate
consequence of \autoref{th:rm-confluence}. In the sequel, a notation
for referring to such forms will be useful.

\begin{defn}[Reading and merging normal form]
The notation $|t|$ denotes the $\onereadmerge$-normal form of a
suspension expression $t$.
\end{defn}

It is easily seen that the $\onereadmerge$-normal form for a term 
that does not contain meta variables is a a term that is devoid of
suspensions, \ie, a de Bruijn term. A further observation is that if
the all the environments appearing in the original term are simple,
then just the reading rules suffice in reducing it to the de Bruijn
term that is its unique $\onereadmerge$-normal form.

\section{Simulation of Beta Reduction}
\label{sec:simulation}

A fundamental property of all explicit substitution calculi is
that they properly simulate the lambda calculus. In particular,
we must ensure that our $\beta_s$ rule corresponds to the $\beta$
rule of the lambda calculus, modulo the reading and merging
rules. The following two theorems establish this result. The first
shows that a $\beta_s$ rewrite on suspension terms corresponds to
some number of $\beta$ rewrites in the lambda calculus. One
might think of this theorem as proving the soundness of our
calculus. The second theorem shows that any $\beta$ rewrite can be
simulated using the rules of our calculus.  One might think of
this theorem as proving the completeness of our calculus.

\begin{theorem}
If $x_1$ and $x_2$ are suspension expressions such that $x_1
\onebetas x_2$ then $\rnf{x_1} \mbeta \rnf{x_2}$.
\end{theorem}
\begin{proof}
This result is proven for the original suspension calculus in
\cite{nadathur-98-suspension}. We know from \autoref{thm:same-nf} that
$\onereadmerge$-normal forms of $x_1$ and $x_2$ are the same as in the
original suspension calculus, thus we can carry over the previous
result.
\end{proof}

\begin{theorem}
If $x_1$ and $x_2$ are suspension expressions in
$\onereadmerge$-normal form such that $x_1 \onebeta x_2$ then $x_1
\all x_2$.
\end{theorem}
\begin{proof}
A stronger version of this property, where the result is replaced with
$x_1 \readbetas x_2$, is proved as Lemma 8.2 of the original
suspension paper. Since the $\beta_s$ and the reading rules are
essentially the same for the two calculi, the result carries over.
\end{proof}

\section{Confluence of Overall Calculus}
\label{sec:confluence-all}

In this section we prove that the full suspension calculus is
confluent even in the presence of graftable meta variables.  One
important distinction between the proof of this property and the proof
of confluence for the reading and merging rules is that the latter is
proven in spite of the merging rules while the former will be proven
only because of the merging rules (see \autoref{sec:susp-metavars}
for a discussion of why the merging rules are required). Thus this
property speaks to the well designed nature of the merging
rules. Moreover, this property makes the suspension calculus a
candidate for unification procedures designed specifically for
graftable meta variables \cite{dowek-95-higherorder} because with this
confluence property we are guaranteed that $\oneall$-normal forms are
unique even with graftable meta variables.

For most explicit substitution calculi, confluence of the full
calculus is proven using Hardin's Interpretation Method
\cite{hardin-89-confluence}.  The interpretation method starts with
the lambda calculus which is already confluent and it uses the
confluence and strong termination of the substitution rules (in our
case, the reading and merging rules) to close a confluence diagram for
the overall calculus. This method is inadequate, however, when we
allow for graftable meta variables. The problem is that the lambda
calculus with graftable meta variables does not make sense and is not
confluent. Instead, we follow the method presented in
\cite{curien-96-confluence} which is based the on the following key
lemma.
\begin{lemma}
\label{lem:RSR}
Let $\mathcal{R}$ and $\mathcal{S}$ be two
relations defined on the same set $X$, $\mathcal{R}$ being
confluent and strongly normalizing, and $\mathcal{S}$ being
strongly confluent, i.e. such that the following diagrams hold for
any $f,g,h\in X$:
\begin{center}
\begin{minipage}{5cm}
\centerline{
\xymatrix{
f \ar@{->}[dd]_{\mathcal{S}} \ar@{->}[rr]^{\mathcal{S}} &&
g \ar@{-->}[dd]^{\mathcal{S}} \\
\\
h \ar@{-->}[rr]^{\mathcal{S}} && k \\
}}
\end{minipage}
\hspace{1in}
\begin{minipage}{5cm}
\centerline{
\xymatrix{
f \ar@{->}[dd]_{\mathcal{R}} \ar@{->}[rr]^{\mathcal{S}} &&
g \ar@{-->}[dd]^{\mathcal{R}^*} \\
\\
h \ar@{-->}[rr]^{\mathcal{R^*SR^*}} && k \\
}}
\end{minipage}
\end{center}
Then the relation $\mathcal{R}^*\mathcal{S}\mathcal{R}^*$ is
confluent.
\end{lemma}

We will apply the lemma using the reading and merging rules as
$\mathcal{R}$ and parallel $\beta_s$-reduction as $\mathcal{S}$.

\begin{defn}[Parallel $\beta_s$-reduction]
Parallel $\beta_s$-reduction is defined by the rules in
\autoref{fig:parallel} and is denoted by $\one{\beta_s||}$.
\end{defn}

\begin{figure}[t]
\begin{align*}
&\infer{t \to t}{}
&
&\infer{e \to e}{} \\
\\
&\infer{t_1\app t_2 \to t_1'\app t_2'}{t_1 \to t_1' && t_2 \to t_2'}
&
&\infer{(t,l) :: e \to (t',l) :: e'}{t \to t' && e \to e'} \\
\\
&\infer{\lambdadb t \to \lambdadb t'}{t \to t'}
&
&\infer{\menv{e_1,nl_1,ol_2,e_2} \to \menv{e_1',nl_1,ol_2,e_2'}}{
  e_1 \to e_1' && e_2 \to e_2'} \\
\\
&\infer{\env{t,ol,nl,e} \to \env{t',ol,nl,e'}}{t \to t' && e \to e'}
&
&\infer{(t,n) \to (t',n)}{t \to t'} \\
\\
&\infer{(\lambdadb t_1)\app t_2 \to \env{t_1', 1, 0, (t_2',0)::nil}}{
  t_1 \to t_1' && t_2 \to t_2'}
\end{align*}
\caption{Parallel $\beta_s$-reduction}
\label{fig:parallel}
\end{figure}

\begin{lemma}
$\onereadmerge$ and $\one{\beta_s||}$ satisfy the condition of
\autoref{lem:RSR}.
\end{lemma}
\begin{proof}
$\one{\beta_s||}$ is obviously strongly confluent since $\onebetas$ is
is a left linear system with no critical pairs. This proves that the
first diagram in \autoref{lem:RSR} holds.

For the second diagram, the interesting case is the critical pair
for $f = \env{(\lambdadb t_1)\app t_2, ol, nl, e}$. In this case, we
have $g = \env{\env{t_1', 1, 0, (t_2',0)::nil}, ol, nl, e'}$ and $h =
\env{\lambdadb t_1, ol, nl, e}\app \env{t_2, ol, nl, e}$, where $t_1
\one{\beta_s||} t_1'$, $t_2 \one{\beta_s||} t_2'$, and $e
\one{\beta_s||} e'$. We must find a $k$ such that $g \readmerge k$ and
$h \readmerge h' \one{\beta_s||} h'' \readmerge k$. This is
straightforward since
\begin{align*}
g &= \env{\env{t_1', 1, 0, (t_2',0)::nil}, ol, nl, e'} \\
&\onemerge \env{t_1', ol+1, nl, \menv{(t_2',0)::nil, 0, ol, e'}} \\
&\readmerge \env{t_1', ol+1, nl, (\env{t_2', ol, nl, e'}, nl)::e'}
\end{align*}
and
\begin{align*}
h &= \env{\lambdadb t_1, ol, nl, e}\app \env{t_2, ol, nl, e} \\
&\oneread \lambdadb \env{t_1, ol+1, nl+1, (\#1,nl+1)::e}\app \env{t_2,
  ol, nl, e} \\
&\one{\beta_s||} \env{\env{t_1', ol+1, nl+1, (\#1,nl+1)::e'}, 1, 0,
  (\env{t_2', ol, nl, e'}, 0)::nil} \\
&\onemerge \env{t_1', ol+1, nl, \menv{(\#1,nl+1)::e', nl+1, 1,
    (\env{t_2', ol, nl, e'}, 0)::nil}} \\
&\readmerge \env{t_1', ol+1, nl, (\env{t_2', ol, nl, e'}, nl)::e'}
\end{align*}
\end{proof}

\begin{theorem}
\label{thm:full-confluence}
The relation $\oneall$ is confluent.
\end{theorem}
\begin{proof}
Note that $\oneall \subseteq \mathcal{R}^*\mathcal{S}\mathcal{R}^*
\subseteq \all$.
\end{proof}

\section{Similarity in the Suspension Calculus}
\label{sec:similarity}

The purpose of this section is to introduce a notion of similarity in
the suspension calculus which relates suspension expressions that
differ only in the renumbering and indices of environment terms. This
allows us to formally capture the notion that two environments are
similar enough that they act the same during rewriting which will be
useful when we translate from another explicit substitution calculus
into the suspension calculus (\autoref{sec:lambdasigma-to-susp}). The
notion of similarity stems from the fact that there are two ways to
represent the renumbering to be done on an environment term. One is
using the difference between the new embedding level of the suspension
and the embedding level of the environment term. The other is with an
explicit renumbering substitution applied to the term in the
environment term. This section proves that these two notions are
equivalent for the purpose of finding normal forms.

\begin{defn}
\label{def:similar}
The similarity relation $\sim$ is defined in \autoref{fig:similar}.
\begin{figure}[t]
\begin{align*}
&\infer{t \sim t}{}
&
&\infer{e \sim e}{} \\
\\
&\infer{t_1\app t_2 \sim t_1'\app t_2'}{t_1 \sim t_1' && t_2 \sim t_2'}
&
&\infer{(t,n)::e \sim (t',n)::e'}{t \sim t' && e \sim e'} \\
\\
&\infer{\lambdadb t \sim \lambdadb t'}{t \sim t'}
&
&\infer{\menv{e_1,nl_1,ol_2,e_2} \sim \menv{e_1',nl_1,ol_2,e_2'}}{
  e_1 \sim e_1' && e_2 \sim e_2'} \\
\\
&\infer{\env{t,ol,nl,e} \sim \env{t',ol,nl,e'}}{t \sim t' && e \sim e'}
&
&\infer{(t,n) \sim (t',n)}{t \sim t'}
\end{align*}
\begin{align*}
\infer{(\env{t,ol,nl,r}, nl + k)::e \sim (\env{t',ol,nl',r'}, nl'
  + k)::e'}{t \sim t' && r \sim r' && e \sim e'}
\end{align*}
\caption{The similarity relation, $\sim$}
\label{fig:similar}
\end{figure}
\end{defn}

The main result of this section is to prove that similar terms rewrite
to the same normal form. This first requires proving the following
lemma.

\begin{lemma}
\label{lem:backwards-m5}
Let $\menv{e_1,nl_1,ol_2,e_2} \readmerge e$ where $e$ is a simple
environment.
\begin{itemize}
\item If $nl_1 - lev(e_1) \geq k$ and $ol_2 \geq k$, then
$\menv{e_1,nl_1-k,ol_2-k,e_2\{k\}} \readmerge e$.
\item If $nl_1 - lev(e_1) \geq ol_2$, then $e_1 \readmerge e$.
\end{itemize}
\end{lemma}
\begin{proof}
The proof is by induction on the length of the sequence
$\menv{e_1,nl_1,ol_2,e_2} \readmerge e$.
\end{proof}

\begin{theorem}
If $t \sim t'$ for terms $t$ and $t'$ then they rewrite by reading and
merging rules to the same de Bruijn term. If $e \sim e'$ for
environments $e$ and $e'$ then they rewrite by reading and merging
rules to similar simple environments.
\end{theorem}
\begin{proof}
We prove the general case of $exp \sim exp'$ for suspension
expressions $exp$ and $exp'$. We do this by induction using the
relation $\gg$ defined in \autoref{def:exprorder}. Note that this
relation decreases when an expression is rewritten using the reading
and merging rules and also a subpart is always smaller than the
original expression.

Because we are inducting using $\gg$, we can assume that the result
already holds for any similar subparts of $exp$ and $exp'$. Then we can
rewrite these similar subparts to be equal in the case of terms or
simple environments in the case of environments. This decreases the
measure of the overall terms $exp$ and $exp'$ and thus the inductive
hypothesis then applies to them. By this reasoning, we can assume that
whenever two subparts of $exp$ and $exp'$ are similar they are in fact
equal in the case of terms and simple in the case of environments. We
will also use the convention that $x$ and $x'$ are always similar.

Let us consider cases based on the structure of $exp$ and $exp'$,
first looking at the case when both are terms. If they are both
constants or de Bruijn indices then the result is trivial. If $exp =
(t_1\app t_2)$ then $exp' = (t_1'\app t_2')$ and $t_1 = t_1'$ and $t_2
= t_2'$ so the result follows trivially. A similar result holds in the
case where $exp$ and $exp'$ are lambda abstractions.

The first nontrivial case is when $exp$ and $exp'$ are both
suspensions, say $exp = \env{t,ol,nl,e}$ and $exp' =
\env{t,ol,nl,e'}$. Now consider which rewrite rules apply to the
toplevel of these terms, keeping in mind that $t$ is in normal
form. If $t$ is an application or an abstraction then (r5) or (r6)
applies and the result follows from the inductive hypothesis. If $t$
is a de Bruijn term and (r2) or (r4) applies then the result again
follows from the inductive hypothesis. If (r3) applies and $e$ and
$e'$ have the same head then the result is trivial. The key case is
when (r3) applies and $e$ and $e'$ have different heads, in which case
we have,
\begin{align*}
exp
&= \env{\#1,ol,nl,(\env{t_r,ol_r,nl_r,r}, nl_r + k)::e_1}\\
&\one{(r3)} \env{\env{t_r,ol_r,nl_r,r}, 0, nl - (nl_r + k), nil}\\
&\one{(m1)} \env{t_r, ol_r, nl - (nl_r + k) + nl_r, \menv{r, nl_r, 0,
    nil}}\\
&\one{(m2)} \env{t_r, ol_r, nl - k, r}\\
\\
exp'
&= \env{\#1,ol,nl,(\env{t_r,ol_r,nl_r',r'}, nl_r' +
  k)::e_1'}\\
&\one{(r3)} \env{\env{t_r,ol_r,nl_r',r'}, 0, nl - (nl_r' + k), nil}\\
&\one{(m1)} \env{t_r, ol_r, nl - (nl_r' + k) + nl_r', \menv{r', nl_r',
    0, nil}}\\
&\one{(m2)} \env{t_r, ol_r, nl - k, r'}
\end{align*}
The two resulting suspensions are similar and smaller than their
original terms, thus the inductive hypothesis finishes this case.

The other half of the proof is to show that when $exp$ and $exp'$ are
similar environments then they rewrite to similar simple
environments. The cases when $exp$ and $exp'$ are either $nil$ or a
cons follow trivially from the inductive hypothesis. The important
case is when $exp = \menv{e_1,nl_1,ol_2,e_2}$ and $exp' =
\menv{e_1',nl_1,ol_2,e_2'}$. Consider the cases for which rewrites can
apply to the toplevel of both expressions. If (m2), (m3), or (m4)
applies to the first expression then the same rewrite applies to the
second environment and the result follows easily. The case when (m5)
applies to both is also direct using the inductive hypothesis. The two
remaining cases are the most interesting: when (m6) applies to both,
and when (m5) applies to one and (m6) to the other.

Consider when (m6) applies to both expressions. Here the head terms of
$e_1$ and $e_1'$ must be the same. If the head terms of $e_2$ and
$e_2'$ are also the same then the result is trivial. Otherwise we
have,
\begin{align*}
exp
&= \menv{(t_1,nl_1)::e_3,nl_1,ol_2,(\env{t_r,ol_r,nl_r,r},
  nl_r+k)::e_4}\\
&\one{(m6)} (\env{t_1,ol_2,nl_r+k,(\env{t_r,ol_r,nl_r,r},
  nl_r+k)::e_4}, nl_r+k+(\monus{nl_1}{ol_2}))\\
&\hspace{1in}::\menv{e_3,nl_1,ol_2,(\env{t_r,ol_r,nl_r,r}, nl_r+k)::e_4}\\
\\
exp'
&= \menv{(t_1,nl_1)::e_3',nl_1,ol_2,(\env{t_r',ol_r,nl_r',r'},
  nl_r'+k)::e_4'}\\
&\one{(m6)} (\env{t_1,ol_2,nl_r'+k,(\env{t_r',ol_r,nl_r',r'},
  nl_r'+k)::e_4'}, nl_r'+k+(\monus{nl_1}{ol_2}))\\
&\hspace{1in}::\menv{e_3',nl_1,ol_2,(\env{t_r,ol_r,nl_r',r'},
  nl_r'+k)::e_4'}\\
\end{align*}
These two environments are still similar and the inductive hypothesis
now applies.

The final case is when (m5) applies to one expression and (m6) to the
other. Without loss of generality, assume that (m6) applies to $exp$
and (m5) to $exp'$. There are two subcases based on whether the heads
of $e_2$ and $e_2'$ are the same or not. Here we will only consider
hardest case where the heads differ. The other case is a
simplification of the following argument,
\begin{align*}
exp
&= \menv{(\env{t_s,ol_s,nl_s,s},nl_s+k_s)::e_3,nl_1,ol_2,
         (\env{t_r,ol_r,nl_r,r},nl_r+k_r)::e_4}\\
&\one{(m6)} (\env{\env{t_s,ol_s,nl_s,s}, ol_2, nl_r+k_r,e_2},
         nl_r+k_r+(\monus{nl_1}{ol_2})) :: \menv{e_3,nl_1,ol_2,e_2}\\
&\one{(m1)} (\env{t_s,ol_s+(\monus{ol_2}{nl_s}),
         nl_r+k_r+(\monus{nl_s}{ol_2}), \menv{s,nl_s,ol_2,e_2}},\\
&\hspace{1in} nl_r+k_r+(\monus{nl_1}{ol_2})) ::
                  \menv{e_3,nl_1,ol_2,e_2}\\
\\
exp'
&= \menv{(\env{t_s',ol_s,nl_s',s'},nl_s'+k_s)::e_3',nl_1,ol_2, e_2'}
\end{align*}
Consider first the case when $nl_1-(nl_s'+k_s) \geq ol_2$.
Since (m6) applies to the first expression we know that $nl_1 = nl_s +
k_s$ and thus $nl_s - nl_s' \geq ol_2$. The first expression then
rewrites to
\begin{multline*}
(\env{t_s,ol_s,nl_r+k_r+nl_s-ol_2, \menv{s,nl_s,ol_2,e_2}},\\
nl_r+k_r+nl_1-ol_2))
:: \menv{e_3,nl_1,ol_2,e_2}
\end{multline*}
The second expression rewrites using (m5) multiple times to
\[ (\env{t_s',ol_s,nl_s',s'},nl_s'+k_s)::e_3' \] It is easily seen
that $exp \gg \menv{s,nl_s,ol_2,e_2}$ and $exp' \gg
\menv{s',nl_s,ol_2,e_2'}$. Moreover, $\menv{s,nl_s,ol_2,e_2}$ and
$\menv{s',nl_s,ol_2,e_2'}$ are similar so the inductive hypothesis
applies and tells us these merged environments rewrite to similar
simple environments. Since $nl_s - ol_2 \geq nl_s' \geq lev(s')$,
applying \autoref{lem:backwards-m5} yields that
$\menv{s,nl_s,ol_2,e_2}$ and $s'$ also rewrite to similar simple
environments. By applying these rewrites, the heads of our two
environments are now similar. By the inductive hypothesis we also know
that $\menv{e_3,nl_1,ol_2,e_2}$ and $\menv{e_3',nl_1,ol_2,e_2'}$
rewrite to similar simple environments. Since $nl_1 - lev(e_3') \geq
ol_2$ we can apply \autoref{lem:backwards-m5} to know that
$\menv{e_3,nl_1,ol_2,e_2}$ and $e_3'$ rewrite to similar simple
environments, thus finishing this case.

The other case is when $nl_1-(nl_s'+k_s) < ol_2$. Then we can apply
(m5) multiple times to $exp'$ and then eventually (m6),
\begin{align*}
exp'
&= \menv{(\env{t_s',ol_s,nl_s',s'},nl_s'+k_s)::e_3',nl_1,ol_2, e_2'}\\
&\many{(m5)} \lmenv (\env{t_s',ol_s,nl_s',s'},nl_s'+k_s)::e_3',
 nl_s'+k_s, ol_2-nl_s+nl_s',\\
&\hspace{1in} e_2'\{nl_s-nl_s'\} \rmenv\\
&\one{(m6)} (\env{\env{t_s',ol_s,nl_s',s'}, ol_2-nl_s+nl_s',
 l, e_2'\{nl_s-nl_s'\}},\\
&\hspace{1in}l + (\monus{k_s}{(ol_2-nl_s)}))::\\
&\hspace{2in} \menv{e_3', nl_s'+k_s, ol_2-nl_s+nl_s',
  e_2'\{nl_s-nl_s'\}}
\end{align*}
Where $l = lev(e_2'\{nl_s-nl_s'\})$. Let us focus first on the tail
portions of our two environments. By the inductive hypothesis
$\menv{e_3,nl_1,ol_2,e_2}$ and $\menv{e_3',nl_1,ol_2,e_2'}$ rewrite to
similar simple environments. Then by applying \autoref{lem:backwards-m5},
$\menv{e_3,nl_1,ol_2,e_2}$ and
$\menv{e_3',nl_s'+k_s,ol_2-nl_s+nl_s',e_2'\{nl_1-nl_s'\}}$ rewrite
to similar simple environments.

Finally we focus on the heads of our environments. The head for
$exp'$ can now be rewritten using (m1),
\begin{align*}
&\env{\env{t_s',ol_s,nl_s',s'}, ol_2-nl_s+nl_s',
 l, e_2'\{nl_s-nl_s'\}}\\
&\one{(m1)} \env{t_s', ol_s + ol_2 - nl_s, l, \menv{s', nl_s',
    ol_2-nl_s+nl_s', e_2'\{nl_s-nl_s'\}}}
\end{align*}
Note that as before $exp \gg \menv{s,nl_s,ol_2,e_2}$ and $exp' \gg
\menv{s',nl_s,ol_2,e_2'}$, also $\menv{s,nl_s,ol_2,e_2}$ and
$\menv{s',nl_s,ol_2,e_2'}$ are similar, so by the inductive hypothesis
these merged environments rewrite to similar simple environments.
Since $nl_s-lev(s')\geq nl_s-nl_s'$ and $ol_2 \geq nl_s-nl_s'$,
applying \autoref{lem:backwards-m5} yields that
$\menv{s,nl_s,ol_2,e_2}$ and
$\menv{s',nl_s',ol_2-nl_s+nl_s',e_2'\{nl_s-nl_s'\}}$ rewrite to
similar simple environments. Using this rewriting results in the heads
being similar.
\end{proof}


\chapter{Comparison with Other Explicit Substitution
  Calculi}
\label{ch:compare}

There are many explicit substitution calculi that offer alternative
treatments of substitutions which leads to varying computational
properties. In this chapter we outline three computational properties
that are of particular interest and use these to categorize various
calculi based on how well they capture these. The three properties we
focus on are (1) combination of substitution walks, (2) confluence in
the presence of graftable meta variables, and (3) preservation of
strong normalization.

Combination of substitution walks, also called merging, can be traced
back to de Bruijn \cite{debruijn-72-nameless}. The substitution
operation on de Bruijn terms, see \autoref{def:debruijn-substitution},
is denoted by $S(t;s_1,s_2,\ldots)$ and represents the term $t$ where
$s_i$ is substituted for the the $i^{th}$ de Bruijn index. De Bruijn
establishes the meta-property $S(S(t;s_1,s_2,\ldots);r_1,r_2,\ldots) =
S(t,u_1,u_2,\ldots)$ where $u_i = S(s_i,r_1,r_2,\ldots)$. Here two
substitutions walks over $t$ are merged into a single substitution
walk over $t$. For a more concrete example, consider the term
$((\lambdadb \lambdadb t_1)\app t_2\app t_3)$. A naive reduction of
this term would require two walks over the structure of $t_1$: the
first for $t_2$ and the second for $t_3$. Moreover, the second walk
would also have to walk over the structure of $t_2$ each place where
it is substituted into $t_1$. A more reasonable approach is to merge
the two substitution prior to making a walk over the structure of
$t_1$. For instance, in the suspension calculus we can rewrite the
term to $\env{t_1,2,0,(t_2,0)::(t_3,0)::nil}$ which requires only one
walk over $t_1$ and avoids any walks over $t_2$ since the
non-overlapping nature of the two substitutions is detected by the
merging process.  In practice, this property has proven to have a
great impact on efficiency \cite{liang-04-choices}.

Confluence in the presence of graftable meta variables, see
\autoref{sec:susp-metavars}, requires a calculus with rules
interactions between substitutions. To see why, consider the example
from \autoref{sec:susp-metavars} in a named context where we have the
term $((\lambda a . ((\lambda b .  X)\app t_1))\app t_2)$ with $X$ a
graftable meta variable. Depending on which redex is contracted first,
this term can reduce to either $X\langle t_1/b\rangle \langle t_2/a
\rangle$ or $X\langle t_2/a \rangle \langle t_1\langle t_2/a\rangle /
b \rangle$ where $\langle t/x\rangle$ is an explicit representation of
substitution. In order to reconcile these two terms, interaction rules
for substitutions must be added to the calculus.  These interaction
rules either take the form of combination rules, as seen in the
previous paragraph, or permutation rules.
  
Preservation of strong normalization (PSN) means that lambda terms
which are strongly normalizing in the lambda calculus remain strongly
normalizing in the explicit substitution calculus, \ie, an infinite
reduction path is never added to a term with only finite reduction
paths. To see why PSN might fail, consider the above problem of
confluence and how we might resolve it by adding a permutation rule of
the form $X\langle t_1/b\rangle \langle t_2/a \rangle \ra X\langle
t_2/a \rangle \langle t_1\langle t_2/a\rangle / b \rangle$. This rule
fixes the confluence problem, but now the system fails to preserve
strong normalization since the new rule can be repeatedly applied to
itself to permute the substitutions back and forth. Preservation of
strong normalization is desirable because one often works in a
representational setting with typed lambda terms which are strongly
normalizing and if PSN holds then all of those terms remain strongly
normalizing in the explicit substitution calculus. On the other hand,
if PSN does not hold then one must be careful in selecting a reduction
strategy which avoids the newly introduced infinite reduction paths.
Preservation of strong normalization is studied in further depth in
\autoref{ch:psn}.

Another part of our survey of explicit substitution calculi consists
of translations between the other popular calculi and the suspension
calculus towards understanding and contrasting their relative
capabilities. To give substance to our translations, we established
relevant properties of the translations such as their correctness and
their ability to preserve important computational properties of the
calculi they relate.  The first half of showing correctness is that
well-formed terms are translated to well-formed terms. The second half
is that normal forms, with respect to substitutions, are preserved by
the translation.  To show that important properties are preserved, we
will argue that the translations are information preserving, which is
an intuitive, rather than formal notion.  There are various ways in
which we can capture this notion with the most desirable being to show
that if $t$ rewrites to $r$ in one step then given the translation
$T$, $T(t)$ rewrites to $T(r)$ in at least one step. We call this
property {\it simulation} because it shows the translation preserves
the information needed to simulate the substitution process of one
calculus using the substitution process of another. This is not always
possible due to the idiosyncrasies of different calculi. In these
cases we will look we will find other ways of arguing for information
preservation, while also looking at why simulation fails since the
reason often reveals key differences between calculi.  Finally, we
note that for all of our translations we assume we are in a context
without graftable meta variables.

We begin by separating the calculi based on combination of
substitution walks, since this property is evident and has the
greatest effect on the syntax of the language. 

\section{Calculi Without Merging}

In this section we look at three calculi without merging:
$\lambda\upsilon$-calculus \cite{benaissa-96-upsilon}, $\lambda
s$-calculus \cite{kamareddine-95-calculus}, and $\lambda s_e$-calculus
\cite{kamareddine-97-extending}. These calculi lack the syntax for
merging substitutions and instead each substitution in these calculi
represents a substitution for at most one de Bruijn index and then
possible renumberings for other de Bruijn indices. Since the notion of
substitution in the suspension calculus is more general, we can only
discuss a translation from these calculi to the suspension calculus
and not the other way around. Nevertheless, seeing how the
substitution concepts in these calculi are reflected in the suspension
calculus gives us greater insight into their key characteristics.

\subsection{The \texorpdfstring{$\lambda\upsilon$}{Lambda
    Upsilon}-calculus}
\label{sec:lambdaupsilon}

The $\lambda\upsilon$-calculus is actually a simplification of the
$\lambda\sigma$-calculus, a calculus we will see more of in
\autoref{sec:lambdasigma}. The $\lambda\upsilon$-calculus was created
by removing the syntax for merging of substitutions available in the
$\lambda\sigma$-calculus and then modifying the rewriting rules to
accommodate the new syntax. This simplified system was proven to
preserve strong normalization, but at the cost of confluence in the
presence of graftable meta variables. Another cost of the
simplification is a peculiarity in the rewriting rules that make the
system undesirable from an implementation perspective. We develop
these issues in this section, starting first with the syntax of the
calculus.

\begin{defn}
The syntax of $\lambda\upsilon$-expressions is given by the following
definitions of terms, denoted $a$ and $b$, and substitutions, denoted
$s$.
\begin{tabbing}
\qquad\=$\langle ET\rangle$\=\quad::=\quad\=\kill
\>$a$\>\quad::=\quad\>$\underline{n} \ \vert \ a\app b\ \vert\
\lambdadb a\ \vert\ a[s]$\\
\>$s$\> \quad::=\quad \> $a/\ \vert \lift(s)\ \vert\
\uparrow$
\end{tabbing}
\end{defn}

The term $\underline{n}$ represents the $n^{th}$ de Bruijn index,
and $a[s]$ is called a {\it closure}. The substitution
$a/$ is called {\it slash} and represents the substitution of $a$
for the first de Bruijn index and a shifting down of all other de
Bruijn indices. The substitution $\lift(s)$ is called {\it
lift} and is used to push substitutions underneath lambda
abstractions. The last substitution $\uparrow$ is called {\it
shift} and represents increasing all free de Bruijn indices by
one. Many of these concepts are less generalized versions of what is
available in the suspension calculus, a point we make more explicit
with the following translation.

\begin{defn}\label{def:lambda-upsilon-to-susp}
The translation $T$ from $\lambda\upsilon$-terms to suspension terms
and the translation $E$ from $\lambda\upsilon$-substitutions to
triples of an old embedding level, a new embedding level, and a
suspension environment are defined simultaneously by recursion as
follows:
\begin{enumerate}
\item For a term $t$, $T(t)$ is $\#n$ if $t$ is $\underline{n}$,
  $(T(a)\app T(b))$ if $t$ is $(a\app b)$, $\lambdadb T(a)$ if $t$ is
  $\lambdadb a$, and $\env{T(a),ol,nl,e}$ if $t$ is $a[s]$ where
  $(ol,nl,e) = E(s)$.

\item For a substitution $s$, $E(s)$ is $(1,0,(T(a),0)::nil)$ if $s$
  is $a/$, $(0,1,nil)$ if $s$ is $\uparrow$, and
  $(ol+1,nl+1,(\#1,nl+1)::e)$ if $s$ is $\lift(s')$ where $(ol,nl,e) =
  E(s')$.
\end{enumerate}
\end{defn}

\begin{theorem}
For every $\lambda\upsilon$-term $a$, $T(a)$ is a well-formed
suspension term.
\end{theorem}
\begin{proof}
The proof is by induction using the dual property that for every
$\lambda\upsilon$-substitution $s$ such that $(ol,nl,e) = E(s)$ we
have $ol = len(e)$, $nl \geq lev(e)$, and $e$ is a well-formed
suspension environment.
\end{proof}

\begin{figure}[t]
\begin{tabbing}
\qquad \=(Lambda)\qquad\= $(\lambdadb a)[s] \ra \lambdadb a[\lift(s)]$
\qquad\qquad\= (FVarLift)\qquad\=\kill

\>(B)\> $(\lambdadb a)\app b \ra a[b/]$
\>(VarShift)\> $\underline{n}[\uparrow] \ra \underline{n+1}$\\[10pt]

\>(App)\> $(a\app b)[s] \ra a[s]\app b[s]$
\>(FVarLift)\> $\underline{1}[\lift(s)] \ra \underline{1}$ \\[3pt]
\>(Lambda)\> $(\lambdadb a)[s] \ra \lambdadb a[\lift(s)]$
\>(RVarLift)\> $\underline{n+1}[\lift(s)] \ra
                    \underline{n}[s][\uparrow]$\\[10pt]

\>(FVar)\> $\underline{1}[a/] \ra a$ \\[3pt]
\>(RVar)\> $\underline{n+1}[a/] \ra \underline{n}$
\end{tabbing}
\caption{Rewrite rules for the $\lambda\upsilon$-calculus}
\label{fig:lambda-upsilon-rules}
\end{figure}

The rules of the $\lambda\upsilon$-calculus are presented in
\autoref{fig:lambda-upsilon-rules}. We define the $\upsilon$ rules to
be all the rules of the $\lambda\upsilon$-calculus except (B). Because
there is no possibility for merging substitutions, the $\upsilon$
rules simply push substitutions down in the tree and then evaluate
them once they are applied to de Bruijn indices. Thus most of the
$\upsilon$ rules can be matched up with a corresponding reading rule
from the suspension calculus, with the exception being the rule
RVarLift. The fundamental problem with this rule is that it replaces a
single substitution on the left with two substitutions on the right.
From the suspension calculus point of view, this is a step backwards.
Thus we instead prove the following theorem in which $a \one{\upsilon}
b$ implies $T(a)$ and $T(b)$ rewrite to a common term rather than a
stronger one in which $T(a) \readmerge T(b)$.

\begin{theorem}
Let $a$ and $b$ be $\lambda\upsilon$-terms such that $a
\one{\upsilon} b$. Then there exists a suspension term $t$ such that
$T(a) \readmerge t$ and $T(b) \readmerge t$.
\end{theorem}
\begin{proof}
The proof is by case analysis on the rule used to transition from $a$
to $b$. In every case but RVarLift we can actually prove that $T(a)
\readmerge T(b)$. The most difficult of these cases is FVar for which
we must show $\env{\#1, 1, 0, (T(a),0)::nil} \readmerge T(a)$. To do
this, we first apply (r3) to generate $\env{T(a),0,0,nil}$. Then we
prove by induction the general property that $\env{t,0,0,nil}
\readmerge t$ in a setting without graftable meta variables.

In the case of RVarLift, suppose that $(ol,nl,e) = E(s)$. Then we can show
that the terms $\env{\#(n+1),ol+1,nl+1,(\#1,nl+1)::e}$ and
$\env{\env{\#n,ol,nl,e},0,1,nil}$ have a common reduct in the term
$\env{\#n,ol,nl+1,e}$.
\end{proof}

Based on this theorem, the translation $T$ preserves de Bruijn normal
forms. To show that $T$ is information preserving we offer the
following theorem which shows that $T$ is one-to-one.

\begin{theorem}
The translation $T$ is one-to-one.
\end{theorem}
\begin{proof}
The proof is by induction using the dual property that $E$ is
one-to-one.
\end{proof}

Looking again at the RVarLift rule, we can see a problem from the
implementation perspective. Consider the term
$\underline{4}[\lift(\lift(\lift(a/)))]$ which rewrites to
$a[\uparrow][\uparrow][\uparrow]$. Here three separate renumbering
passes are generated in order to increase all free de Bruijn indices
by three. The problem is that not only is combination of substitutions
not allowed, but the syntax of the $\lambda\upsilon$-calculus is not
rich enough to encode a renumbering of de Bruijn indices by anything
but one. In the next section we will see another calculus without
merging, but with a more general notion of substitution which avoids
this problem.

\subsection{The \texorpdfstring{$\lambda s$}{lambda s}-calculus}
\label{sec:lambdas}

The $\lambda s$-calculus is similar to the $\lambda\upsilon$-calculus
in that it preserves strong normalization and fails to have confluence
in the presence of graftable meta variables. The two primary
differences are that the $\lambda s$-calculus clearly separates the
processes of substitution and renumbering, and the $\lambda
s$-calculus has more general notion of substitution. These differences
are reflected in the syntax.

\begin{defn}
The syntax of $\lambda s$-expressions is given by the following
definition of terms, denoted $a$ and $b$.
\begin{tabbing}
\qquad\=$\langle ET\rangle$\=\quad::=\quad\=\kill
\>$a$\>\quad::=\quad\>${\tt n} \ \vert \ a\app b\ \vert\
\lambdadb a\ \vert\ a \sig{i} b\ \vert\ \ph{k}{i} a$
\end{tabbing}
Here ${\tt n}$ and $i$ range over all positive integers and $k$ over all
non-negative integers.
\end{defn}

The term ${\tt n}$ represents the $n^{th}$ de Bruijn index. The term
$a \sig{i} b$ is called a {\it closure} and represents the
substitution of a renumbered version of $b$ for the $i^{th}$ de Bruijn
index in $a$ and a shifting down by one of all de Bruijn indices
greater than $i$ in $a$. The term $\ph{k}{i} a$ is called an {\it
  update} and represents an increase by $i-1$ of all de Bruijn
indices greater than $k$. All of these concepts can be translated
into the suspension calculus by the following translation.

\begin{defn}\label{def:lambda-s-to-susp}
The translation $T$ from $\lambda s$-terms to suspension terms is
defined by recursion as follows: For a term $t$, $T(t)$ is $\#n$ if
$t$ is ${\tt n}$, $(T(a)\app T(b))$ if $t$ is $(a\app b)$, $\lambdadb
T(a)$ if $t$ is $\lambdadb a$, $\env{T(a), i, i-1, (\#1,i-1) ::
  (\#1,i-2) :: \ldots :: (\#1, 1) :: (T(b),0)::nil}$ if $t$ is $a
\sig{i} b$, and $\env{T(a), k, k+i-1, (\#1,k+i-1) :: (\#1,k+i-2) ::
  \ldots :: (\#1,i) :: nil}$ if $t$ is $\ph{k}{i} a$.
\end{defn}

\begin{theorem}
For every $\lambda s$-term $a$, $T(a)$ is a well-formed suspension
term.
\end{theorem}
\begin{proof}
The proof is by induction.
\end{proof}

\begin{figure}[t]
\begin{tabbing}
\qquad \=$\phi-app-transition$\qquad\=\kill
\>$\sigma$-$generation$\> $(\lambdadb a)\app b \ra a \sig{1} b$ \\[3pt]

\>$\sigma$-$\lambda$-$transition$\>
    $(\lambdadb a) \sig{i} b \ra \lambdadb (a \sig{i+1} b)$\\[3pt]

\>$\sigma$-$app$-$transition$\>
    $(a_1\app a_2) \sig{i} b \ra (a_1 \sig{i} b)\app (a_2 \sig{i} b)$\\[3pt]

\>$\sigma$-$destruction$\>
    ${\tt n} \sig{i} b \ra
    \begin{cases}
      {\tt n - 1} & \text{if $n > i$} \\
      \ph{0}{i} b & \text{if $n = i$} \\
      {\tt n}     & \text{if $n < i$}
    \end{cases}$\\[3pt]

\>$\varphi$-$\lambda$-$transition$\>
    $\ph{k}{i} (\lambdadb a) \ra \lambdadb (\ph{k+1}{i} a)$\\[3pt]

\>$\varphi$-$app$-$transition$\>
    $\ph{k}{i} (a_1\app a_2) \ra (\ph{k}{i} a_1)\app (\ph{k}{i} a_2)$\\[3pt]

\>$\varphi$-$destruction$\>
    $\ph{k}{i} {\tt n} \ra
    \begin{cases}
      {\tt n + i - 1} & \text{if $n > k$} \\
      {\tt n}     & \text{if $n \leq k$}
    \end{cases}$

\end{tabbing}
\caption{Rewrite rules for the $\lambda s$-calculus}
\label{fig:lambda-s-rules}
\end{figure}

The rules of the $\lambda s$-calculus are presented in
\autoref{fig:lambda-s-rules}.  We define the $s$ rules to be all the
rules of the $\lambda s$-calculus except $\sigma$-$generation$.
Because of the separation between substitution and renumbering, there
is some redundancy in the rules, \eg\ $\sigma$-$app$-$transition$ and
$\varphi$-$app$-$transition$. But looking at
$\sigma$-$\lambda$-$transition$ and $\varphi$-$\lambda$-$transition$,
we see the benefit is that substitution and renumbering can have
separate behaviors for descending underneath lambda abstractions. This
cleanness in the rules allows for very well behaved translation.

\begin{theorem}\label{thm:step-lambda-s}
Let $a$ and $b$ be $\lambda s$-terms such that $a \one{s} b$. Then
$T(a) \posread T(b)$.
\end{theorem}
\begin{proof}
The proof is by case analysis on the rule used to transition from $a$
to $b$.
\end{proof}

The above theorem tells us that the translation $T$ is information and
normal form preserving. Moreover, it shows us that the suspension
calculus, even without merging, is capable of exactly simulating the
$\lambda s$-calculus, and it gives us a proof that the $s$ rules of
the $\lambda s$-calculus are strongly normalizing since the reading
(and merging) rules are strongly normalizing. In the original paper on
the $\lambda s$-calculus a similar translation is proven from the $\lambda
s$-calculus to the $\lambda\sigma$-calculus, and it is through this
translation that the strong normalization of the $s$ rules is established.
That the $\lambda s$-calculus translates so nicely into both the suspension
calculus and the $\lambda\sigma$-calculus is a strong argument that
the calculus is well-designed and natural for representing single
substitutions.

Another point to note about the above theorem is that the suspension
calculus may have to make multiple reading steps to simulate a single
step in the $\lambda s$-calculus. The primary reason for this comes
from the (r3) and (r4) rules of the suspension calculus which are used
to evaluate the result of applying a suspension to a de Bruijn
index. Because a suspension can representing a substitution for
various different de Bruijn indices, these rules must check to see
which substitution applies for a given index. In the $\lambda
s$-calculus on the other hand, each closure represents a single
substitution so when a de Bruijn index is encountered we can
immediately check if it is the one being substituted for. This clearly
gives a benefit in efficiency to the $\lambda s$-calculus, but this
benefit is not enough to offset the benefit gaining by merging
substitutions \cite{liang-04-choices}.

\subsection{The \texorpdfstring{$\lambda s_e$}{Lambda se}-calculus}
\label{sec:lambdase}

The $\lambda s_e$-calculus is an extension of the $\lambda s$-calculus
in order to gain confluence in a setting with graftable meta
variables. The calculus achieves this by allowing what some call
merging of substitutions, but what is more accurately described as
permutation of substitutions. To be precise, the $\lambda
s_e$-calculus maintains the same syntax as the $\lambda s$-calculus
and extends the rewrite rules with the six rules in
\autoref{fig:lambda-se-rules}.

\begin{figure}
\begin{tabbing}
\qquad\=$\varphi$-$\varphi$-$transition$ 2\qquad\=
$(a \sig{i} b)\sig{j} c \ra (a \sig{j+1} c) \sig{i} (b \sig{j-i+1}
c)$\qquad\=\kill

\>$\sigma$-$\sigma$-$transition$
\>$(a \sig{i} b)\sig{j} c \ra
     (a \sig{j+1} c) \sig{i} (b \sig{j-i+1} c)$
\>if $i \leq j$\\[3pt]

\>$\sigma$-$\varphi$-$transition$ 1
\>$(\ph{k}{i} a)\sig{j} b \ra \ph{k}{i-1} a$
\>if $k < j < k + i$\\[3pt]

\>$\sigma$-$\varphi$-$transition$ 2
\>$(\ph{k}{i} a)\sig{j} b \ra \ph{k}{i}(a \sig{j-i+1} b)$
\>if $k + i \leq j$\\[3pt]

\>$\varphi$-$\sigma$-$transition$
\>$\ph{k}{i}(a\sig{j} b) \ra (\ph{k+1}{i} a)\sig{j} (\ph{k+1-j}{i} b)$
\>if $j \leq k + 1$\\[3pt]

\>$\varphi$-$\varphi$-$transition$ 1
\>$\ph{k}{i}(\ph{l}{j} a) \ra \ph{l}{j}(\ph{k+1-j}{i} a)$
\>if $l + j \leq k$\\[3pt]

\>$\varphi$-$\varphi$-$transition$ 2
\>$\ph{k}{i}(\ph{l}{j} a) \ra \ph{l}{j+i-1} a$
\>if $l \leq k < l + j$
\end{tabbing}
\caption{Additional rewrite rules for the $\lambda s_e$-calculus}
\label{fig:lambda-se-rules}
\end{figure}

Notice that each rule has careful restrictions on it to prevent
looping behavior, but as shown in \cite{guillaume-00-preserve} this is
not enough: the $\lambda s_e$-calculus fails to preserve strong
normalization. Another more technical problem with the $\lambda
s_e$-calculus is that normal forms when in a context of graftable meta
variables can become unwieldy. A $\lambda s_e$-normal form has the
same basic structure as a de Bruijn term, except that graftable meta
variables can have sequences of closures and updates applied to them.
The only restriction on these substitutions is that none of the
$s_e$-rules apply to them \cite{kamareddine-97-extending}. This
problem is brought to the forefront in the context of higher-order
unification using the $\lambda s_e$-calculus, where graftable meta
variables and normal forms play an important role in efficient
unification procedures \cite{rincon-03-applying}.

\section{A Calculus with Merging: the
  \texorpdfstring{$\lambda\sigma$}{Lambda Sigma}-calculus}
\label{sec:lambdasigma}

The $\lambda\sigma$-calculus supports a general notion of composition
of substitution walks and there exists a variant of it which is
confluent in a setting with graftable meta variables
\cite{abadi-91-explicit, curien-96-confluence}. Unfortunately,
Mellies was able to demonstrate that the calculus lacks preservation
of strong normalization by presenting a simply typed lambda term for
which an infinite $\lambda\sigma$-reduction path exists
\cite{mellies-95-terminate}.

We use the rest of this section to define the $\lambda\sigma$-calculus
and construct translations to and from the suspension calculus.

\begin{defn}
The syntax of $\lambda\sigma$-expressions is given by the following
definition of terms, denoted $a$ and $b$, and substitutions, denoted
$s$ and $t$.
\begin{tabbing}
\qquad\=$\langle ET\rangle$\=\quad::=\quad\=\kill
\>$a$\>\quad::=\quad\>$1 \ \vert \ ab\ \vert\ \lambda a\ \vert\
a[s]$\\
\>$s$\> \quad::=\quad \> $id\ \vert \uparrow \vert\ a \cdot s\
\vert\ s \circ t$
\end{tabbing}
\end{defn}

The term $a[s]$ is called a {\it closure} and represents the term $a$
with some substitution $s$ to be applied to it. The substitution $id$
is the identity substitution. The substitution $\uparrow$ is called
{\it shift} and represents a increasing of all free de Bruijn indices
by 1. The substitution $a \cdot s$ is called {\it cons} and represents
a term $a$ to be substituted for the first de Bruijn index along with
a substitution $s$ for the remaining indices. Lastly, the substitution
$s \circ t$ represents the merging of the substitution $s$ and $t$.

Note that the terms in this calculus only contain the first de Bruijn
index. All others are represented by $1[\uparrow]$, $1[\uparrow \circ
\uparrow]$, $1[(\uparrow \circ \uparrow)\ \circ \uparrow]$ etc. We
abbreviate the $n^{th}$ de Bruijn index as $1[\uparrow^{n-1}]$. With
this in mind, the rules for the $\lambda\sigma$-calculus are presented
in \autoref{fig:lambda-sigma-rules}.

\begin{figure}[t]
\begin{tabbing}
\qquad \=(Abs)\qquad\= $(\lambda a)[s] \ra \lambda a[1\cdot (s\ \circ
\uparrow)]$ \qquad\qquad\= (ShiftCons)\qquad\= $\uparrow \circ\
(a \cdot s) \ra s$ \kill
\>(Beta)\> $(\lambda a) b \ra a[b\cdot id]$ \\[20pt]

\>(App)\> $(a\ b)[s] \ra a[s]\ b[s]$ 
\>(VarId)\> $1[id] \ra 1$ \\[3pt]
\>(Abs)\> $(\lambda a)[s] \ra \lambda a[1\cdot (s\ \circ
\uparrow)]$
\>(VarCons)\> $1[a\cdot s] \ra a$ \\[3pt]
\>(Clos)\> $a[s][t] \ra a[s\circ t]$
\>(IdL)\> $id\circ s \ra s$ \\[3pt]
\>(Map)\> $(a\cdot s)\circ t \ra a[t]\cdot (s\circ t)$
\>(ShiftId)\> $\uparrow \circ\ id \ra\ \uparrow$ \\[3pt]
\>(Ass)\> $(s\circ t)\circ u \ra s \circ (t\circ u)$
\>(ShiftCons)\> $\uparrow \circ\ (a \cdot s) \ra s$
\end{tabbing}
\caption{Rewrite rules for the $\lambda\sigma$-calculus}
\label{fig:lambda-sigma-rules}
\end{figure}

\subsection{Suspension Expressions to
  \texorpdfstring{$\lambda\sigma$}{Lambda Sigma}-expressions}

The translation from suspension expressions to
$\lambda\sigma$-expressions works by unfolding the information which
is represented by the indices and embedding levels of the suspension
calculus into individual substitution operations of the
$\lambda\sigma$-calculus. Accounting for this, the rest of the
translation is straightforward and translates suspension expressions
into corresponding $\lambda\sigma$-expressions: suspension to closure,
nil to id, cons to cons, and merged to merged. Besides the difference
in representing renumberings, the syntax of the two calculi match up
nicely.

\begin{defn}\label{def:susp-to-lambda-sigma}
The translation $S$ from suspension terms to $\lambda\sigma$-terms and the
translation $R$ from pairs of a suspension environment and a new embedding
level to $\lambda\sigma$-substitutions are defined simultaneously by
recursion as follows:
\begin{enumerate}
\item For a term $t$, $S(t)$ is $1$ if $t$ is $\#1$, $1[\uparrow^n]$
  if $t$ is $\#(n+1)$ with $n \geq 1$, $(S(a)\app S(b))$ if $t$ is
  $(a\app b)$, $\lambdadb S(a)$ if $t$ is $\lambdadb a$, and
  $S(t')[R(e,nl)]$ if $t$ is $\env{t,ol,nl,e}$.

  \item For an environment $e$ and natural number $j$, $R(e,j)$ is
  $(\ldots (id\ \overbrace{\circ \uparrow)\ \circ \uparrow) \circ
    \ldots}^j)$ if $e$ is $nil$, $(\ldots ((S(t) \cdot R(e',n))\
  \overbrace{\circ \uparrow)\ \circ \uparrow) \circ \ldots}^{j-n})$ if
  $e$ is $(t,n)::e'$, and $R(e_1, nl_1) \circ R(e_2, j -
  (\monus{nl_1}{ol_2}))$ if $e$ is $\menv{e_1, nl_1, ol_2, e_2}$.
\end{enumerate}
\end{defn}

The translation from suspension expressions to
$\lambda\sigma$-expressions includes a translation $R(e, j)$ which
translates the environment $e$ relative to the embedding level $j$.
Restrictions must be placed on this translation to ensure the
definition is well-formed. For example, looking at the second case for
$R(e,j)$, one might worry that $j < n$ in which case having $(j-n)$
shifts does not make sense. We can ensure this never happens by
requiring that $lev(e) \leq j$ every time $R(e,j)$ is called.
Enforcement of this is provided by the wellformedness properties of
suspension terms.

\begin{theorem}
If $t$ is a suspension term then $S(t)$ is well-defined.
\end{theorem}
\begin{proof}
The property must be proved simultaneously with the property that if $e$ is
a suspension environment and $j$ is an integer with $j \geq lev(e)$
then $R(e,j)$ well-defined.
\end{proof}

Due to occurrences of the identity substitution and small differences
in associativity, the $\lambda\sigma$-calculus does not simulate the
suspension calculus. Instead, we show that normal forms are preserved
by the translation.

\begin{theorem}
Let $a$ and $b$ be suspension terms such that $a \onereadmerge
b$. Then there exists a $\lambda\sigma$-term $t$ such that $S(a)
\many{\sigma} t$ and $S(b) \many{\sigma} t$.
\end{theorem}
\begin{proof}
The proof uses the dual property that if $a$ and $b$ are suspension
environments and $j$ is an integer such that $j \geq lev(a)$ then
$R(a,j) \readmerge R(b,j)$. We can then prove both properties by case
analysis on the rule used to transition from $a$ to $b$.
\end{proof}

Finally, we argue that $S$ is information preserving by showing that
it is one-to-one. Note that this property is quite strong since we are
translating from a calculus with merging to another calculus with
merging. Before, when we translated from a calculus without merging to
one with merging, this property was more obvious since only the
substitution representations might overlap. Here we must worry about
substitution representations and also merged substitution
representations.

\begin{theorem}
The translation $S$ is one-to-one.
\end{theorem}
\begin{proof}
The result follows easily from noticing that $R(e,j)$ can never equal
$\uparrow^k$ for any $e$, $j$, and $k$.
\end{proof}

Because the $\lambda\sigma$-calculus and the suspension calculus seem
to be equally expressive we can define a translation in the other
direction in the following section.

\subsection{\texorpdfstring{$\lambda\sigma$}{Lambda Sigma}-expressions
  to Suspension Expressions}
\label{sec:lambdasigma-to-susp}

The translation from $\lambda\sigma$-expressions to suspension expressions
proceeds in the obvious way except for a special case when translating
$s\ \circ \uparrow$ which changes the shift substitution into the
corresponding renumbering concept expressed in embedding levels and
indices.

\begin{defn}\label{def:lambda-sigma-to-susp}
The translation $T$ from $\lambda\sigma$-terms to suspension terms and
the translation $E$ from $\lambda\sigma$-substitutions to triples of
an old embedding level, a new embedding level, and a suspension
environment are defined simultaneously by recursion as follows:
\begin{enumerate}
\item For a term $t$, $T(t)$ is $\#1$ if $t$ is $1$, $\#(n+1)$ if $t$
  is $1[\uparrow^n]$, $(T(a)\app T(b))$ if $t$ is $(a\app b)$,
  $\lambdadb T(a)$ if $t$ is $\lambdadb a$, and $\env{T(a),ol,nl,e}$ if
  $t$ is $a[s]$ where $(ol,nl,e) = E(s)$.

\item For a substitution $s$, $E(s)$ is $(0,0,nil)$ if $s$ is $id$,
  $(0,1,nil)$ if $s$ is $\uparrow$,\\ $(ol+1, nl, (T(a),nl)::e)$ if
  $s$ is $a\cdot s'$ where $(ol,nl,e) = E(s')$, $(ol,nl+1,e)$ if $s$
  is $s'\ \circ \uparrow$ where $(ol,nl,e) = E(s')$, and $(ol_1 +
  (\monus{ol_2}{nl_1}), nl_2 + (\monus{nl_1}{ol_2}),
  \menv{e_1,nl_1,ol_2,e_2})$ if $s$ is $s_1 \circ s_2$ where
  $(ol_1,nl_1,e_1) = E(s_1)$ and $(ol_2,nl_2,e_2) = E(s_2)$.
\end{enumerate}
If more than one case apply to the same expression, we require that
first one listed is the one used.
\end{defn}

\begin{theorem}
For every $\lambda\sigma$-term $a$, $T(a)$ is a well-formed
suspension term.
\end{theorem}
\begin{proof}
The proof is by induction using the dual property that for every
$\lambda\sigma$-substitution $s$ such that $(ol,nl,e) = E(s)$ we
have $ol = len(e)$, $nl \geq lev(e)$, and $e$ is a well-formed
suspension environment.
\end{proof}

The suspension calculus is not capable of simulating the
$\lambda\sigma$-calculus and this is not a bad property. If the
suspension calculus were able to simulate the
$\lambda\sigma$-calculus, then the suspension calculus would be able
to simulate the Mellies counterexample to preservation of strong
normalization \cite{mellies-95-terminate}. The primary reason for the
lack of simulation is the Ass rule which establishes an association
rule for merged substitutions. In the suspension calculus, however,
such association is a property we have proven with significant work,
see \autoref{sec:assoc}, but it is not a rule of the calculus.
Instead of showing simulation, we show normal forms are preserved by
the translation.

\begin{theorem}
Let $a$ and $b$ be $\lambda\sigma$-terms such that $a \one{\sigma}
b$. Then there exists a suspension-term $t$ such that $T(a)
\readmerge t$ and $T(b) \readmerge t$.
\end{theorem}
\begin{proof}
A naive approach to this theorem would be filled with special cases to
account for the special cases present in the translations $T$ and $E$.
In order to avoid this note that in the special case of
$T(1[\uparrow^n]) = \#(n+1)$ if we had used the more general
translation of $a[s]$ we would have produced $\env{\#1, 0, n, nil}$.
Since these two terms are $\readmerge$-convertible we can pick the
second one for this theorem and ignore the special case. The same
result holds for the special case of $E(s\ \circ \uparrow)$.

The other difficulty in proving this theorem is that we will need a
corresponding property for $\lambda\sigma$-substitutions. Naively,
this property might be that if $s \one{\sigma} t$ then the old and new
embedding level components of $E(s)$ and $E(t)$ are equal and the
environment components rewrite to a common environment. This will fail
because of the (Map) rule in the $\lambda\sigma$-calculus which has
the form $(a\cdot s_1)\circ s_2 \ra a[s_2]\cdot (s_1\circ s_2)$.
Letting $t = T(a)$, $(ol_1,nl_1,e_1) = E(s_1)$, and $(ol_2,nl_2,e_2) =
E(s_2)$, the environment components of the translation $E$ applied to
the left and right sides of the (Map) rule are
$\menv{(t,nl_1)::e_1,nl_1,ol_2,e_2}$ and
$(\env{t,ol_2,nl_2,e_2},nl_2+(\monus{nl_1}{ol_2}))::\menv{e_1,nl_1,ol_2,e_2}$,
respectively. Note that this is very similar to our rule (m6) but
different in that $e_2$ might not have the form $(s,l)::e_2'$ and also
we use $nl_2$ instead of the level of $e_2$. Because of these
problems, these two environments are not rewritable to a common
environment. Instead, we generalize the property to state that the
environment components should be rewritable to similar environments,
see \autoref{sec:similarity}, from which the result follows.
\end{proof}

The translation $T$ is not one-to-one because the
$\lambda\sigma$-substitutions $\uparrow$ and $id\ \circ \uparrow$
translate to the same tuple. We can, however, prove that this
translation is a left-inverse of the translation $S$ from suspension
term to $\lambda\sigma$-terms.  Because the translation $S$ is
information preserving, this result is strong evidence that $T$ is
also information preserving. Moreover, this result shows that our
translations are well balanced and therefore, hopefully natural.

\begin{theorem}
For every suspension term $t$, $T(S(t)) = t$.
\end{theorem}
\begin{proof}
The proof is by induction using the dual property that for every
suspension environment $e$ and integer $j$ such that $j \geq lev(e)$,
we have $E(R(e,j)) = (ol, j, e)$ where $ol = len(e)$.
\end{proof}


\chapter{Preservation of Strong Normalization}
\label{ch:psn}

Normal forms hold a special place in the lambda calculus. The normal
form of a term has the same meaning as the original term without any
$\beta$-contraction work left. This makes normal forms an ideal basis
for unification procedures over the lambda calculus. By focusing on
normal forms, such procedures can ignore the $\beta$-contraction
aspect of the lambda calculus, and instead focus just on the binding
structure of normal forms.

Because of this lofty position, much work has been put into
determining when normal forms exist and how to compute them. For
instance, a strong motivation behind the simply typed lambda calculus
in \autoref{sec:typed} is that normal forms are guaranteed to exist
for all terms in the calculus. Such terms are called {\it
  normalizable}. In fact, any reduction of a term from that calculus
is finite and must reach the normal form, prompting the title {\it
  strongly normalizable}. This is not always the case for other
variants of the lambda calculus. In those instances, a reduction
strategy must be carefully chosen which reduces a term to its normal
form, provided one exists. A well known strategy which has this
behavior is called {\it normal order} reduction and consists of always
contracting the leftmost outermost $\beta$-redex, which we will call
the {\it leading redex}.  The fundamental reason why this strategy
works is that contracting anything other than leading redex will not
affect the existence of the leading redex. Thus the leading redex will
persist forever unless we contract it, and therefore we choose to
contract it first. In this chapter we look at generalizations of these
notions of normalizability, strong normalizability, and reduction
strategies to the context of explicit substitution calculi.

\section{Preservation of Normalizability}

Explicit substitution calculi are elaborations of the lambda calculus
and as such they preserve normalizability. That is, given a
normalizable term in the lambda calculus, it remains normalizable in
an explicit substitution calculus. The reason is is that each
$\beta$-contraction step in the lambda calculus can be matched in an
explicit substitution calculus by a simulated $\beta$-contraction step
followed by a series of substitution steps.  This approach to
computing normal forms is logically sound, but it removes the
practical benefits of using an explicit substitution calculus. We can
make two improvements on it.

The first improvement is actually one which can be realized in the
lambda calculus. Because we are often interested in finding normal
forms for the purpose of unification, we can develop a weaker notion
of normal form which does not require as much work to compute, but
still suffices for the purpose of unification. Such a notion is
capture by reducing terms to the form $(\lambdadb \lambdadb \ldots\
\lambdadb (h\app t_1\app \ldots\app t_n))$ where $h$ is either a
constant, a de Bruijn index, or a meta variables. This is called a
{\it head normal form}. Another term in head normal form then unifies
with this one if and only if it has the form $(\lambdadb \lambdadb
\ldots\ \lambdadb (h\app s_1\app \ldots\app s_n))$ where the number of
leading lambdas is the same and $t_i$ unifies with $s_i$ for $1 \leq i
\leq n$. This method saves us from having to normalize the terms $t_1,
\ldots, t_n, s_1, \ldots, s_n$ in the case that unification fails. A
simple and complete procedure for computing head normal forms is to
perform normal order reduction until a head normal form is reach and
then stop there. This is called {\it head normalization} and in this
case the leading redex is referred to as the {\it head redex}.

The second improvement we can make is to generalize the notion of head
normalization to the explicit substitution context. Nadathur has made
such a generalization in the case of the suspension calculus and has
proven that such a procedure always finds the normal form of a
normalizable term \cite{nadathur-99-finegrained}. The idea behind
Nadathur's notion of head reduction is to define a head redex as
either the head $\beta$-redex or a $\onereadmerge$-redex which occurs
above the head $\beta$-redex. Generalized head reduction then consists
of contracting head redexes until no more head reductions are
possible. This generalization provides more freedom in computing head
normal forms, but it retains the essential property of head reduction:
that a head normal form is always reached in a finite number of steps,
assuming one exists. To see why this is true, notice that the reading
and merging rules are terminating, so any infinite reduction would
have to consist of infinitely many $(\beta_s)$ applications on head
redexes. But we can map each of these applications to the contraction
of the corresponding $\beta$-redex in the lambda calculus. This is as
simple as taking the $\onereadmerge$-normal form before and after
applying $(\beta_s)$. Thus any infinite generalized head reduction
sequence in the suspension calculus can be mapped onto an infinite
head reduction sequence in the lambda calculus.

\section{Preservation of Strong Normalization}

A more complicated issue is whether a strongly normalizing term in the
lambda calculus remains strongly normalizing within an explicit
substitution calculus. A calculus is said to have {\it preservation of
  strong normalization} (PSN) if this is the case for every strong
normalizing term. This is a desirable property because it speaks to
the coherence of the calculus.  Intuitively one might expect this
property to always hold since explicit substitution calculi are
elaborations of the lambda calculus, but this is not the case. The
interaction of contraction and substitution rules in explicit
substitution calculi creates the possibility of reduction sequences
which do not correspond to reductions in the lambda calculus. The
existence of such reduction sequences speaks poorly for the structure
of an explicit substitution calculus. In the remainder of this
chapter, we focus on the issue of preservation of strong normalization
in various explicit substitution calculi.

\section{PSN in Calculi without Substitution Interaction}

In calculi without rules for interactions between substitutions,
preservation of strong normalization is usually true. In such calculi,
substitutions are generated by a step which simulates
$\beta$-contraction and then those substitutions are pushed down
through the term until they can be evaluated. Because of this
essentially linear path, it is very easy to take an arbitrary
substitution and determine which $\beta$-contraction generated it. By
connecting these $\beta$-contractions in the explicit substitution
calculus to $\beta$-contractions in the lambda calculus, any infinite
reduction sequence in the former can be mapped onto one in the latter.

To take an example, we consider here the proof of preservation of
strong normalization for the $\lambda s$-calculus (see
\autoref{sec:lambdas}). In this setting a closure refers to a term of
the form $a\sig{i} b$ and the inside of a closure refers to the term
$b$. Suppose we have an infinite reduction of some term in this
calculus. We know that the $s$ rules of the $\lambda s$-calculus are
strongly normalizing, so the infinite reduction must contain
infinitely many contractions of $\beta$-redexes using the
$\sigma$-generation rule. At each step of this infinite reduction, we
can look at the $s$-normal form of the current term to see what
progress is being made with respect to the lambda calculus. Clearly
each step which uses an $s$ rule does not change the $s$-normal
form. For the $\sigma$-generation steps, some may change the
$s$-normal form and some may leave it the same. Those that change the
$s$-normal form correspond to $\beta$-contractions in the lambda
calculus. If there are an infinite number of such steps then we can
use the $s$-normal forms as an infinite reduction sequence in the
lambda calculus. The other possibility is that only finitely many
$\sigma$-generation steps correspond to changes in the $s$-normal
form. Now any $\sigma$-generation step which occurs at the top level
(outside of any closures) will be one of these steps which changes the
$s$-normal form, and therefore only finitely many of our
$\sigma$-generation steps occur at the top level. Because there are
only finitely many such steps, we can find a point in our infinite
reduction at which all $\sigma$-generation steps occur inside of
closures. By the infinite pigeonhole principle, there must be one
closure which contains an infinite reduction inside it. Thus we have
reduced our infinite reduction sequence to an infinite reduction which
occurs entirely within a single closure.

The next key step in the proof is to trace each closure back to the
$\beta$-redex from which it was created. This is possible since only
the $\sigma$-generation rule can create closures. Using this idea we
can take our infinite reduction which occurs within some closure, say
$a\sig{i} b$, and know that it came from a term of the form
$((\lambdadb a')\app b')$ where $b'$ rewrites to $b$. Now instead of
contracting the $\beta$-redex we can follow the infinite reduction
path which exists for $b'$. Because this $b'$ is no longer inside of a
closure, the $\sigma$-generation steps inside it will correspond to
reductions in the lambda calculus for the $s$-normal form. By the same
reasoning we have followed so far, what must
occur is that this $b'$ eventually generates a closure which contains
an infinite reduction. But this closure can again be unwound and
mapped into a reduction sequence in the lambda calculus. By repeating
this process indefinitely we generate an infinite reduction sequence in
the lambda calculus.

\section{Problems for Calculi with Substitution Interaction}

Some calculi have rules of interactions between substitutions, the
nature of which depend on the motivation for including them in the
calculus. One motivation for substitution interaction is to regain
confluence in a setting with graftable meta variables. For instance,
in the $\lambda s$-calculus, consider the term $((\lambdadb X)\app
b)\sig{i} c$ where $X$ is a graftable meta variable and $b$ and $c$
are arbitrary terms. On the one hand we can contract the redex to
obtain $(X\sig{1} b)\sig{i} c$, while on the other we can first
distribute the substitution and then perform the reduction to generate
$(X\sig{i+1} c)\sig{1} (b\sig{i} c)$. These two terms cannot be
reduced to a common term because $X$ is a graftable meta variable. In
order to fix this, the $\lambda s_e$-calculus, see
\autoref{sec:lambdase}, extends the $\lambda s$-calculus and
introduces rules for interactions between substitutions
\cite{kamareddine-97-extending}. One of these rules deals exactly with
the case we have,
\begin{tabbing}
\qquad $\sigma$-$\sigma$-$transition$\qquad 
$(a \sig{i} b)\sig{j} c \ra
     (a \sig{j+1} c) \sig{i} (b \sig{j-i+1} c)$\qquad
if $i \leq j$
\end{tabbing}
Applying this rule reconciles the two reductions into the term
$(X\sig{i+1} c)\sig{1} (b\sig{i} c)$.

The danger in admitting a permutation rule is that once we have
permuted two substitutions, we might try to permute them again. The
$\lambda s_e$-calculus tries to avoid this situation by placing side
conditions on the permutation rules so that one substitution can only
be permuted inside another if the outside one represents a contraction
from higher up in the term than the inner substitution. For example,
in a term of the form $((\lambdadb \ldots (\lambdadb a)\app b
\ldots)\app c)$ we allow the contraction of the outer redex to be
permuted inside the contraction of the inner redex. An additional
wrinkle, however, is that we must also add rules for interactions with
updating functions.  It is exactly these additional interaction rules
which causes the $\lambda s_e$-calculus to lose preservation of strong
normalization, as proved by Guillaume~\cite{guillaume-00-preserve}.

Guillaume and David solved this problem by introducing the
$\lambda_{ws}$-calculus which replaces update functions with labels
representing the renumbering to be done
\cite{guillaume-01-weakening}. These labels are then part of the
normal forms of terms in the calculus since there are no rules for
propagating their effects. Thus this calculus corresponds to a version
of the $\lambda s_e$-calculus where restrictions are placed on the
ability to propagate updating functions. These restrictions are not so
severe that confluence in presence of graftable meta variables is
lost.  Furthermore, with these restrictions, Guillaume and David are
able to show that if a term has an infinite reduction sequence then
contracting and propagating a leading redex preserves that infinite
reduction sequence. Using this they map every infinite reduction in
the $\lambda_{ws}$-calculus into an infinite normal order reduction
sequence in the lambda calculus, thus proving preservation of strong
normalization. The key to using the leading redex is that it is above
every other substitution which may be encountered during
propagation. Thus it can be permuted inside of those substitutions
without disturbing their effect on the infinite reduction.

A different approach to substitution interaction is to allow full
combination of substitutions such as in the suspension calculus and
the $\lambda\sigma$-calculus \cite{abadi-91-explicit}. The benefit of
this approach is that the resulting calculus is often very efficient
for direct implementation. Additionally, the merging rules for these
calculi are usually strong enough that confluence in the presence of
graftable meta variables can be recovered. The downside of allowing
merging is that it creates new interaction possibilities for
substitutions, and this may lead to the loss of preservation of strong
normalization. Such is the case in in the $\lambda\sigma$-calculus
where Mellies demonstrated a strongly normalizing term along with an
infinite $\lambda\sigma$-reduction~\cite{mellies-95-terminate}.

The essential problem with merging in the $\lambda\sigma$-calculus is
that superfluous terms can be generated and then allowed to interact
with other substitutions. For instance in a substitution of the form
$\uparrow \circ\ (a \cdot s)$ we know that the term $a$ is going to be
eventually pruned by the shift. However if we have an outer
substitution applied to this substitution, $(\uparrow\circ\ (a\cdot
s)) \circ r$, then we can apply the association rule for merged
environments to rewrite this term to $\uparrow \circ\ ((a\cdot s)\circ
r)$. From here we can map the substitution $r$ onto $a$ to yield
$a[r]$ which is again superfluous, but may contain significant
reduction work.

This idea is played out in completely in the Mellies counterexample
which begins with a term of the form $((\lambdadb a)\app b)[s]$ and
rewrites it so that the substitution $s$ is able to interact with a
version of itself.
\begin{align*}
((\lambdadb a)\app b)[s] &\one{App} (\lambdadb a)[s]\app b[s] \\
&\one{Abs} \lambdadb a[1\cdot (s\ \circ\uparrow)]\app b[s] \\
&\one{Beta} a[1\cdot (s\ \circ\uparrow)][b[s]\cdot id] \\
&\one{Clos} a[(1\cdot (s\ \circ\uparrow))\circ (b[s]\cdot id)] \\
&\one{Map} a[1[b[s]\cdot id] \cdot ((s\ \circ\uparrow) \circ (b[s]\cdot
id))] \\
&\one{Ass} a[1[b[s]\cdot id] \cdot (s\circ (\uparrow \circ\
(b[s]\cdot id)))]
\end{align*}
From here on we can focus solely on the substitution $s \circ
(\uparrow\circ\ (b[s]\cdot id))$. Notice that at this point the $b[s]$
component here is vacuous. Because of the $\uparrow$, the $b[s]$
should be removed as soon as we apply ShiftCons. Unfortunately, the
rules of the $\lambda\sigma$-calculus allow us to play with this
vacuous term and produce an infinite sequence. If we consider that $s$
might be of the form $((\lambdadb a)\app b)\cdot id$ and if we
abbreviate $(\uparrow\circ\ (b[s]\cdot id))$ as $s'$ then we can
rewrite the term $s \circ (\uparrow\circ\ (b[s]\cdot id))$ as follows.
\begin{align*}
s \circ (\uparrow\circ\ (b[s]\cdot id)) &= (((\lambdadb a)\app b)\cdot
id) \circ s' \\
&\one{Map} ((\lambdadb a)\app b)[s'] \cdot (id \circ s') \\
&\one{IdL} ((\lambdadb a)\app b)[s'] \cdot s' \\
&\one{App} ((\lambdadb a)[s']\app b[s']) \cdot s' \\
&\one{Abs} (\lambdadb a[1 \cdot (s'\ \circ\uparrow)]\app b[s'])\cdot
s' \\
&\one{Beta} (a[1\cdot (s'\ \circ\uparrow)][b[s']\cdot id])\cdot s' \\
&\one{Clos} (a[(1\cdot (s'\ \circ\uparrow))\circ (b[s']\cdot id)]\cdot
s' \\
&\one{Map} a[1[b[s']\cdot id] \cdot ((s'\ \circ\uparrow)\circ
(b[s']\cdot id))] \cdot s' \\
&\one{Ass} a[1[b[s']\cdot id] \cdot (s'\circ (\uparrow\circ\
(b[s']\cdot id)))] \cdot s'
\end{align*}
Here we again have a subterm of the form $s'\circ (\uparrow\circ\
(b[s']\cdot id))$. Using this we can repeat the above reasoning to
produce an infinite sequence.

\section{Status of PSN for the Suspension Calculus}

Preservation of strong normalization for the suspension calculus is an
open problem. In this section we explain why the counterexample from
the $\lambda\sigma$-calculus and the proof techniques of the $\lambda
s$-calculus are insufficient in resolving this problem.

To start, consider how the counterexample from the
$\lambda\sigma$-calculus would proceed in the suspension calculus. The
term $((\lambdadb a)\app b)[s]$ in the $\lambda\sigma$-calculus
corresponds to a term $\env{(\lambdadb a)\app b, ol, nl, e}$. Then the
reduction can proceed as follows.
\begin{align*}
\lenv (\lambdadb a)\app b&, ol, nl, e \renv \\
&\one{(r5)} \env{\lambdadb a, ol,
  nl, e}\app \env{b, ol, nl, e} \\
&\one{(r6)} (\lambdadb \env{a, ol+1, nl+1, (\#1,nl+1)::e})\app \env{b,
  ol, nl, e} \\
&\one{(\beta_s)} \env{\env{a, ol+1, nl+1, (\#1,nl+1)::e}, 1, 0,
  (\env{b, ol, nl, e}, 0)::nil} \\
&\one{(m1)} \env{a, ol+1, nl, \menv{(\#1,nl+1)::e, nl+1, 1, (\env{b,
      ol, nl, e}, 0)::nil}} \\
&\one{(m6)} \lenv a, ol+1, nl, (\env{\#1, 1, 0, (\env{b, ol, nl, e},
    0)::nil}, nl)::\\ 
&\hspace{2in}\menv{e, nl+1, 1, (\env{b, ol, nl, e}, 0)::nil}\renv \\
&\one{(m5)} \env{a, ol+1, nl, (\env{\#1, 1, 0, (\env{b, ol, nl, e},
    0)::nil}, nl)::\menv{e, nl, 0, nil}} \\
&\one{(m2)} \env{a, ol+1, nl, (\env{\#1, 1, 0, (\env{b, ol, nl, e},
    0)::nil}, nl)::e}
\end{align*}
At this point we can focus on the first term in the environment and
reduce it as follows.
\begin{align*}
\env{\#1, 1, 0, (\env{b, ol, nl, e}, 0)::nil} &\one{(r3)}
\env{\env{b, ol, nl, e}, 0, 0, nil} \\
&\one{(m1)} \env{b, ol, nl, \menv{e, nl, 0, nil}} \\
&\one{(m2)} \env{b, ol, nl, e}
\end{align*}
This leaves the original term as $\env{a, ol+1, nl, (\env{b, ol, nl,
    e}, nl)::e}$. There doesn't appear to be any means for an infinite
reduction from this since we don't have the environment $e$ acting on
itself, as was the case in the $\lambda\sigma$-calculus. Instead let
us reconsider the steps we took in producing this term and ask if we
could have chosen a different reduction path once we merged the two
environments. The answer is that there can be no other reduction path
for this merged environment in this case or for any merged environment
in the general case. Looking at the rules which operate on merged
environments, we see that there are no choices in which rule can be
applied at a given stage except for the trivial overlap between (m2)
and (m3). Thus the merging process is deterministic. Furthermore,
given an environment of the form $\env{e_1, nl_1, ol_2, e_2}$, there
are no rules which allow the outside context of this environment to
have an effect on $e_1$ or $e_2$, until after the merging is
performed. In this way, the merging process of the suspension calculus
can be viewed as an atomic action. We can conceivably imagine
replacing (m1)-(m6) with a single merging rule which rewrites a term
of the form $\env{\env{t, ol_1, nl_1, e_1}, ol_2, nl_2, e_2}$ to one
of the form $\env{t, ol', nl', e'}$ where $e'$ is a simple
environment. The benefit of the separate rules (m1)-(m6) is that this
large merging operation is done in a lazy fashion.

On the other hand, we can think of extending the proof of preservation
of strong normalization for the $\lambda s$-calculus to apply to the
suspension calculus. In the $\lambda s$-calculus we are able to trace
each closure back to the $\beta$-redex which created it, because only
the $\beta$-contraction rule can generate closures. In the suspension
calculus, environment terms can be created either by
$\beta$-contraction or by merging of substitutions. Thus tracing an
environment term back to a single $\beta$-redex in the lambda calculus
is extremely difficult. Furthermore, the $\lambda s$-calculus had a
fairly linear order in generating and propagating substitutions, while
the suspension calculus has the possibility that propagating one
substitution might mean merging with other substitutions along the
way. This allows the suspension calculus more much freedom in choosing
reduction paths, but it also makes mapping those reduction paths onto
the lambda calculus significantly more difficult.


\chapter{Conclusion}
\label{ch:conc}

In this thesis we have presented a version of the suspension calculus
which combines the desirable theoretical properties of the original
suspension calculus with the practical benefits of the derived
suspension calculi. This new version is created not by adding more to
the calculus, but by simplifying what is already there. This
simplification has the additional benefit of rationalizing the
structure of the calculus, making it possible to easily superimpose
additional logical structure over it. We have illustrated this
capability by showing how typing in the lambda calculus can be treated
in the resulting framework and by presenting a natural translation
into the $\lambda\sigma$-calculus. We have also shown how the
substitution mechanism supports combination of substitution walks
while remaining confluent and terminating. Building on this, we have
proven that the full suspension calculus is confluent even in the
presence of graftable meta variables. The question of preservation of
strong normalization relative to the suspension calculus remains open.
However, we conjecture that it is true and we have presented arguments
as to why this belief might be correct.  If this property is indeed
true then it would make the suspension calculus the only explicit
substitution calculus which possesses the three properties deemed to be
most desirable.

Another contribution of this thesis is a survey of the realm of
explicit substitution calculi. We have utilized the suspension
calculus in this process. In particular, we have described
translations between other popular calculi and the suspension calculus
towards understanding and contrasting their relative capabilities. 
To give substance to this approach, we have established relevant
properties of the translations such as their correctness and their
ability to preserve important computational properties of the calculi
they relate. 

This thesis would be incomplete without a discussion of the possible
ways of building on the results it presents.

\bigskip

\noindent {\bf Preservation of Strong Normalization}\\
Preservation of strong normalization is a problem of significant
theoretical interest because it speaks to the coherence of the
calculus. As already mentioned, we believe that it can be proven true
in the case of the suspension calculus.
The basis of this belief is that the merging (or
permutation) of substitutions which has caused the property to fail in
other calculi is handled correctly in the suspension calculus.
Whereas the $\lambda\sigma$-calculus, the only other calculus that
allows combination of substitutions, allows us to make choices in how
to unravel substitution combination, the suspension calculus treats
substitution combination as a deterministic and pseudo-atomic function.
While this intuition appears to be accurate, working it out into a
proof has been difficult. In particular, reflecting an
arbitrary reduction sequence in the suspension calculus into one in
the lambda calculus appears complicated but this seems to be necessary
to show that infinite sequences in first context must be matched by
infinite ones in the second. Nevertheless, we believe that the simpler
set of combination rules gives us a better handle on this matter
and hence are hopeful of using it to construct an actual argument.

\bigskip

\noindent {\bf Higher-Order Unification using the Suspension
  Calculus}\\ As mentioned in the introduction, one benefit of using
  explicit substitutions is that it allows substitution notions to be
  actively used by processes that operate on the lambda calculus such
  as unification. Recent work has exploited this feature in producing
  a higher-order unification procedures based on a variant of the
  $\lambda\sigma$-calculus which supports graftable meta variables
  \cite{dowek-95-higherorder}.  The original suspension calculus also
  supports graftable meta variables and so this unification idea could
  have been worked out in its context as well. However, the incentive
  for doing this has been small because the complexity of its
  combination rules limits the benefit of doing this in actual
  implementations. Derived calculi based on the suspension calculus
  simplify these combination rules into a couple rules which are
  useful for head normalization, but these calculi are not confluent
  when graftable meta variables are added. By contrast, the suspension
  calculus presented in this thesis has the property of confluence
  even in the presence of graftable meta variables and also has a
  collection of combination rules that is simple enough to use
  directly in an implementation. The benefit of developing the new 
  approach to unification based on the suspension calculus is that it
  treats renumbering in a more efficient manner than the
  $\lambda\sigma$-calculus and so a higher-order unification procedure
  based on the suspension calculus is likely to have better behavior
  in practice. 

\bigskip

\noindent {\bf Compilation of Strong Reduction}\\
Functional programming languages use a notion of reduction where an
expression that has a top level abstraction is treated as a value.
This form of reduction, where it is unnecessary to look underneath
abstractions, is called {\it weak reduction}.  Weak reduction is easily
performed in an interpretive setting by keeping an environment which
tracks variable bindings. It is also possible to compile weak
reduction and the Categorical Abstract Machine which underlies
the Objective Caml programming language provides a framework for doing
exactly this \cite{cousineau-87-cam}. In the representational use of
the lambda calculus it may be necessary to compare underneath lambda
abstraction leading to the need to perform reductions even in such
contexts. This is called {\it strong reduction}. Explicit substitution
calculi provide a basis for realizing strong reduction and in fact an
interpreted approach has been developed based on the suspension
calculus and used in a $\lambda$Prolog implementation
\cite{liang-04-choices}. An approach to using a compilation based
realization of strong reduction has also 
been described in the context of the Coq system
\cite{gregoire-02-compiled}. However, this approach is somewhat ad-hoc
and is based on repeated calls to the reduction machinery underlying
the categorical abstract machine. We believe a uniform compilation
model can be developed using an explicit substitution notation such as
the suspension calculus.


\bibliographystyle{alpha}
\bibliography{thisbib}

\end{document}